\RequirePackage{fix-cm}
\RequirePackage{amsmath}
\documentclass[smallextended]{svjour3}       % onecolumn (second format)
\usepackage{marginnote} % Required for margin notes
\usepackage{wallpaper} % Required to set each page to have a background
\usepackage{amsmath} % Required for equation customization
\usepackage{amssymb} % Required to include mathematical symbols
\usepackage{xcolor} % Required to specify colors by name
\usepackage{verbatim}	% Required to use multirows comments
\usepackage{pdfpages}	% Required to insert pdf file (the cover)
\usepackage{natbib}
\usepackage{graphicx}
\usepackage{tabularx}
\usepackage{multirow}
\usepackage{enumitem}
\usepackage{booktabs}% http://ctan.org/pkg/booktabs
\usepackage[title,titletoc,toc]{appendix}
\usepackage{lmodern}
\usepackage[T1]{fontenc}
\usepackage{blindtext}
\begin{document}

\title{The ARIEL Mission Reference Sample}




\author{Tiziano Zingales	\and
        Giovanna Tinetti	\and 
        Ignazio Pillitteri  \and
        J\'er\'emy Leconte 	\and
        Giuseppina Micela	\and
        Subhajit Sarkar
}


\institute{Tiziano Zingales \at University College London\\
               INAF- Osservatorio Astronomico di Palermo\\
              %\email{fauthor@example.com}           %  \\
           \and
           Giovanna Tinetti \at University College London\\              
           \and                			
          Ignazio Pillitteri \at INAF - Osservatorio Astronomico di Palermo\\
           \and
            J\'er\'emy Leconte \at
           CNRS, Universit\'e de Bordeaux \\
	     Laboratoire d'Astrophysique de Bordeaux 
            \and
            Giuseppina Micela \at INAF - Osservatorio Astronomico di Palermo 
            \and 
            Subhajit Sarkar \at Cardiff University, Cardiff, UK}

\date{Received: date / Accepted: date}

\maketitle

\begin{abstract}
The ARIEL (Atmospheric Remote-sensing Exoplanet Large-survey) mission concept is one of the three M4 mission candidates selected by the European Space Agency (ESA) for a Phase A study, competing for a launch in 2026. 
ARIEL has been designed to study the physical and chemical properties of a large and diverse sample of exoplanets, and through those understand how planets form and evolve in our galaxy.

 Here we describe the assumptions made to estimate an optimal sample of exoplanets -- including both the already known exoplanets and the "expected'' ones yet to be discovered -- observable by ARIEL  and define a realistic mission scenario.  To achieve the mission objectives, the sample should include gaseous and rocky planets with a range of temperatures around stars of different spectral type and metallicity. The current ARIEL design enables the observation of $\sim$1000 planets, covering a broad range of planetary and stellar parameters, during its four year mission lifetime. This nominal list of planets is expected to evolve over the years depending on the new exoplanet discoveries.
\end{abstract}

\section{Introduction}
\subsection{Mission overview}

Today we know over 3500 exoplanets and more than one third are transiting (http://exoplanets.eu/). These  include Earths, super-Earths, Neptunes and Giant planets around a variety of stellar types. The Kepler space mission has discovered alone more than 1000 new transiting exoplanets between 2009 and 2015 and more than 3000 still unconfirmed planetary candidates. 

The number of  known exoplanets is expected to increase in the next decade thanks to current and future space missions (K2, GAIA, TESS, CHEOPS, PLATO) and a long list of ground-based surveys (e.g. HAT-NET, HARPS, WASP, MEarth, NGTS, TRAPPIST, Espresso, Carmenes). They will detect thousands of new transiting exoplanets. 

ARIEL (Atmospheric Remote-sensing Exoplanet Large-survey) is one of the three candidate missions selected by the European Space Agency (ESA) for its next medium-class science mission due for launch in 2026. The goal of the ARIEL mission is to investigate the atmospheres of several hundreds planets orbiting distant stars in order to address the fundamental questions on how planetary systems form and evolve. Key objective of the mission is to find out whether the chemical composition of exoplanetary atmospheres correlate with basic parameters such as the planetary size, density, temperature, and stellar type and metallicity.
During its four-year mission, ARIEL aims at observing a statistically significant sample of exoplanets,  ranging from Jupiter- and Neptune-size down to super-Earth and Earth-size in the visible and the infrared with its meter-class telescope. The analysis of ARIEL spectra and photometric data will allow to extract the chemical fingerprints of gases and condensates in the planets' atmospheres, including the elemental composition for the most favorable targets. It will also enable the study of thermal and scattering properties of the atmosphere as the planet orbit around the star.

The main purpose of this paper is to estimate an optimal  list of targets observable by ARIEL or a similar mission in ten years time and quantify a realistic mission scenario to be completed in 4 year nominal mission lifetime, including the commissioning phase.

To achieve the mission objectives, the sample should include gaseous and rocky planets with a range of temperatures around stars of different spectral type and metallicity.
With this aim, it is necessary to consider both the already known exoplanets and the ``expected'' ones yet to be discovered. 
The data collected by Kepler allow to estimate the occurrence rate of exoplanets according to their size and orbital periods. Using this planetary occurrence rate   and the number density of stars in the Solar neighbourhood, we can estimate the number of exoplanets expected to exist with a particular size, orbital period range and orbiting  a star of a particular spectral type and metallicity.
Here we describe the assumptions made to estimate an optimal sample of exoplanets observable by ARIEL and define the Mission Reference Sample (MRS). It is clear that  this nominal list of planets will change over the years depending on the new exoplanetary discoveries. 

In Section \ref{sec:the_simulation} we explain the method used to estimate the number and the parameters of the planetary systems yet to be discovered. All the potential ARIEL targets will be presented in Section \ref{sec:global_sample}, where we show all the planets that can be observed individually during the mission lifetime, and out of which we want to select the optimal sample. Section \ref{sec:possible_MRS} is dedicated to the selection and description of an ARIEL MRS  fulfilling  the mission requirements, we compare the proposed ARIEL MRS to the sample expected to be discovered by TESS, confirming that TESS could provide a large fraction of the ARIEL targets. A sample including only planets known today is identified. In Section \ref{sec:other_method} we show a possible MRS which maximises the coverage of the planetary and stellar physical parameters.

\subsection{Description of the models}

We use the ESA Radiometric Model \citep{ESA...Rad...Mod} to estimate the performances of the ARIEL mission given the planetary, stellar and orbital characteristics: namely the stellar type and brightness, the planetary size, mass, equilibrium temperature and atmospheric composition, the orbital period and eccentricity. This tool takes into account the mission instrumental parameters and planetary system characteristics to calculate:
\begin{itemize}
\item	The SNR (Signal to Noise Ratio) that can be achieved in a single transit;
\item	The SNR that can be achieved in a single occultation;
\item	The number of transit/occultation revisits necessary to achieve a specified SNR;
\item	The total number and types of targets that can be included in the mission lifetime.
\end{itemize}

In this work, the input planet list for the radiometric model is a combination of known and simulated exoplanets, as detailed in the following sections. We used the instrument parameters of the ARIEL payload as designed during the phase A study.  To increase the efficiency of our simulations we used a Python tool as a wrap of the ESA Radiometric Model, so we could test different mission configurations that fulfil the mission science objectives. The results were validated with ExoSim, a time domain simulator used for the ARIEL space mission, but thanks to its modularity it can be used to study any transit spectroscopy instrument from space or ground. ExoSim has been developed by \citet{2016SPIE.9904E..3RS, 2015EPSC...10..187S, 2015ExA....40..601P} (see App \ref{app:validation}).

\begin{comment}
\begin{figure}[!htbp]
\centering
\includegraphics[scale=0.3]{System_inputs_requirements.png}
\vspace{0.5cm}
\includegraphics[scale=0.3]{System_inputs_yellow_book.png}
\caption{Snapshot of the system parameters used in the ESA Radiometric model. $N_{min}$ and the $X$ factors are shown in App \ref{app:nminx}. Top: parameters specified in the ARIEL MRD. Bottom: parameters used in this paper.}
\label{fig:sys_inp}
\end{figure}
\end{comment}

% In this article, however, we focus mainly on the performances that can be achieved using the  ARIEL instrument design parameters as studied and refined during the Phase A.

\section{Simulations of planetary systems expected to be discovered in the next decade}\label{sec:the_simulation}

\subsection{Star count estimate}
We used the stellar mass function as obtained from the 10-pc RECONS (REsearch Consortium On Nearby Stars) to estimate the number of stars as a function of the K magnitude. We assume mass-luminosity-K magnitude conversions from \citet{1998A&A...337..403B}. The same procedure was adopted by \citet{ESA...EChO...MRS}. The number of main sequence stars with limit K-mag $m_K = 7$ used to infer the number density of stars in the Solar neighbourhood is shown in Tab \ref{tab:star_counts}.

\begin{table}[!htbp]
\centering
\begin{tabular}{|c|c|c|}
  \hline
  \textbf{Mass (M$_\odot$)} & \textbf{Spectral type} & \textbf{N$_*$ (K $< 7$)} \\ \hline
  1.25 - 1.09 &	F6 - F9 & 5646 \\ \hline
  1.09 - 0.87 &	G0 - G8 &	3356 \\ \hline
  0.87 - 0.65 &	K0 - K5 &	1167 \\ \hline
  0.65 - 0.41 &	K7 - M1 &	386 \\ \hline
  0.41 - 0.22 &	M2 - M3 &	81 \\ \hline
  0.22 - 0.10 &	M4 - late M &	28 \\ \hline

\end{tabular}

\caption{Star counts considering different spectral types with limiting magnitude $m_K=7$.\label{tab:star_counts}}
\end{table}
\noindent 
The number density and the number of stars are related through Eq \ref{eq:star_counts}:
\begin{equation}
\rho_{*} = \frac{N_{*}(K < 7)}{\frac{4}{3} \pi d^3}
\label{eq:star_counts}
\end{equation}
\noindent
where the distance $d$ has been calculated in the ARIEL Radiometric Model \citep{ESA...Rad...Mod} using the relation between K magnitude $m_K$ and the distance $d$:
\begin{equation}
m_K = -2.5 \log{\frac{R_{*}^2S_s(\Delta\lambda)}{d^2S_0^K(\Delta\lambda)}}
\label{eq:magk}
\end{equation}
\noindent
In Eq \ref{eq:magk}, $R_*$ is the stellar radius, $S_0^K (\Delta\lambda)$ is the zero point flux for the standard K-band filter profile, $\Delta\lambda$ is the filter band pass given in \citet{2003AJ....126.1090C} and $S_s (\Delta\lambda)$ the stellar flux density evaluated over the same bandwidth. We neglect the interstellar absorption since our stars are at a relatively short distance.

\begin{table}[!htbp]
\centering
\begin{tabular}{|c|c|}
  \hline
  \multicolumn{2}{|c|}{\textbf{Density}} \\
  \hline
   & \textbf{Star / pc$^3$} \\ \hline
  \textbf{$\rho (\text{F6-F9})$} & 0.0039 \\ \hline
  \textbf{$\rho (\text{G0-G8})$} & 0.0044 \\ \hline
  \textbf{$\rho (\text{K0-K5})$} & 0.0049 \\ \hline 
  \textbf{$\rho (\text{K7-M1})$} & 0.0074 \\ \hline
  \textbf{$\rho (\text{M2-M3})$} & 0.0059 \\ \hline
  \textbf{$\rho (\text{M4 - late M})$} & 0.0118 \\ 
  \hline
\end{tabular}
\label{tab:star_density}
\caption{Main sequence star densities considering different spectral types with limiting  magnitude $m_K=7$}
\end{table}

\subsection{Planetary population and occurrence rate}\label{sec:planetary_occurrence_rate}
\begin{figure}[!htbp]
\centering
\includegraphics[scale=0.35]{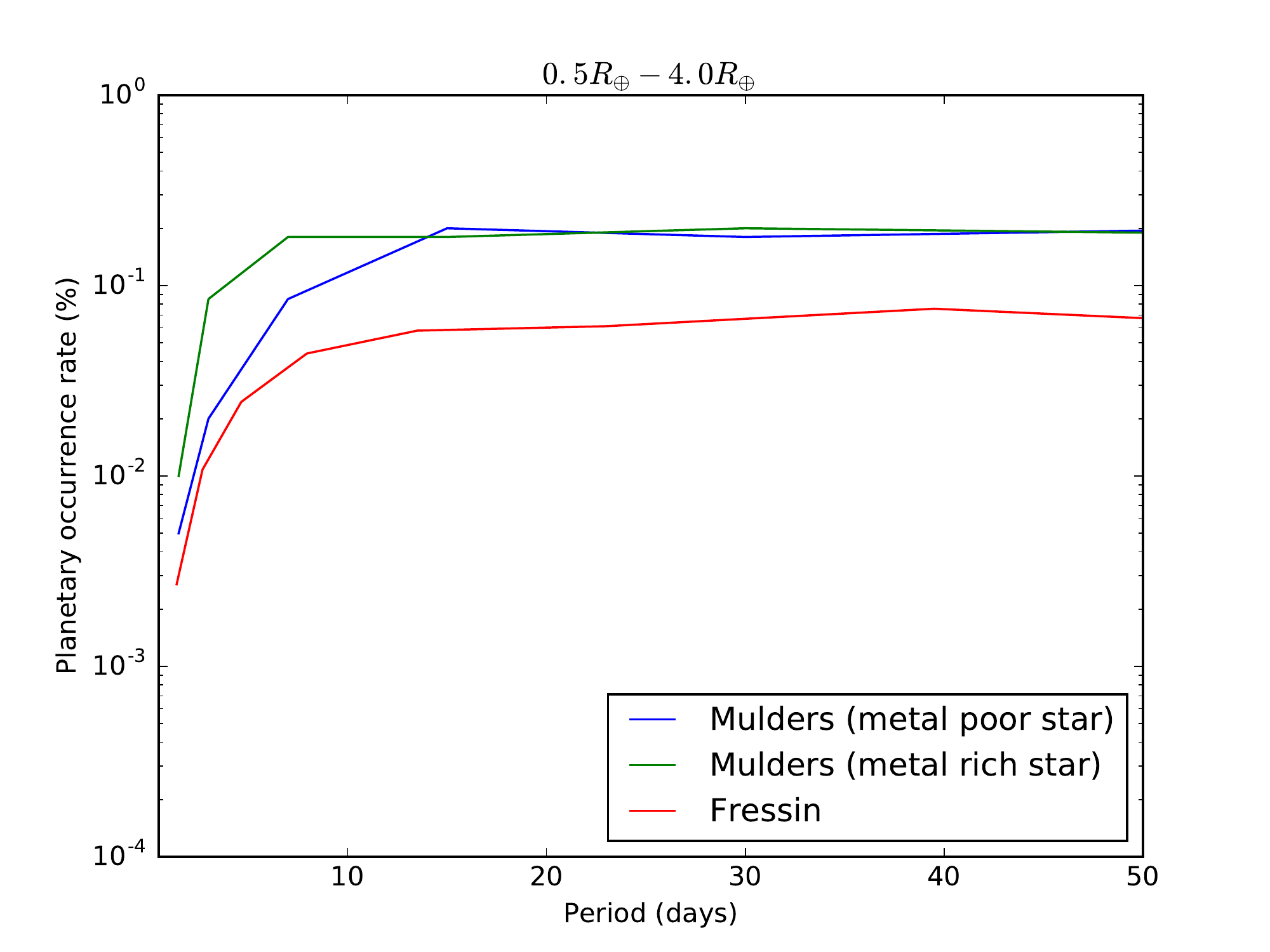}
\caption{Comparison of three different distributions estimating the planetary occurrence rate as a function of orbital period for planets between $0.5 R_{\oplus}$ and $4 R_{\oplus}$. Blue and green lines: results from \citet{2016arXiv160905898M} for two metallicity classes. Red line: results from \citet{2013ApJ...766...81F}. The \citet{2013ApJ...766...81F} statistics strongly underestimates the occurence of sub Neptune size planets compared to \citet{2016arXiv160905898M} and other more recent estimates. The reason is the large number of small planets discovered after 2013.\label{fig:occ_mul}}
\end{figure}

In this section we  briefly review the current knowledge about the occurrence rate of  planets, i.e. the average expected number of planets per star. \citet{2013ApJ...766...81F} used the Kepler statistics to publish the planetary occurrence rates around F, G, K main sequence stars ordered by orbital periods and planetary types. 
An accurate planetary occurrence rate is pivotal to the reliability of the estimate of the existing planets in the Solar neighbourhood.
We used the planetary occurrence rate values for F,G,K and M stars from \citet{2013ApJ...766...81F}, being the most conservative and currently the most complete, i.e. covering all planetary types and stars. Therefore, our estimates for the ARIEL sample are very conservative. 
\citet{2016arXiv160905898M}  updated the planetary occurrence rate for planets between $0.5R_{\oplus}$ and $4R_{\oplus}$ and orbital period $< 50$ days, using a more recent list of planets discovered by the Kepler satellite. Fig \ref{fig:occ_mul} shows the comparison between \citet{2016arXiv160905898M} and \citet{2013ApJ...766...81F}. The differences between the two occurrence rates  can be up to an order of magnitude.  \citet{2015ApJ...814..130M} show that M stars have 3.5 times more small planets ($1.0-2.8 R_{\oplus}$) than  FGK stars, but two times fewer Neptune-sized and larger ($>2.8 R_{\oplus}$) planets. The fraction of M-stars considered in our work is only $\sim 7$\% of the total stellar sample, so we are significantly underestimating the number of small planets around M-dwarfs, which are optimal targets for transit spectroscopy. More recent and complete  results from Mulders and collaborators are expected to be published in the next months and they are not yet available for our simulations. Given the discrepancy between Mulders and Fressin's statistics we expect a substantial improvement in our estimates when the most recent Kepler statistics will become available.

\noindent
\citet{2013ApJ...766...81F} provided the following statistics for  different planetary classes:

\begin{itemize}
\item Jupiters: $6R_{\oplus} < R_p \le 22R_{\oplus}$
\item Neptunes: $4R_{\oplus} < R_p \le 6R_{\oplus}$
\item Small Neptunes: $2R_{\oplus} < R_p \le 4R_{\oplus}$
\item Super Earths: $1.25R_{\oplus} < R_p \le 2R_{\oplus}$
\item Earths: $0.8R_{\oplus} < R_p \le 1.25R_{\oplus}$
\end{itemize}
We adopted a size resolution of $1R_{\oplus}$ in each of these classes.

\begin{figure}[!htbp]
\centering
\includegraphics[scale=0.35, angle=-90]{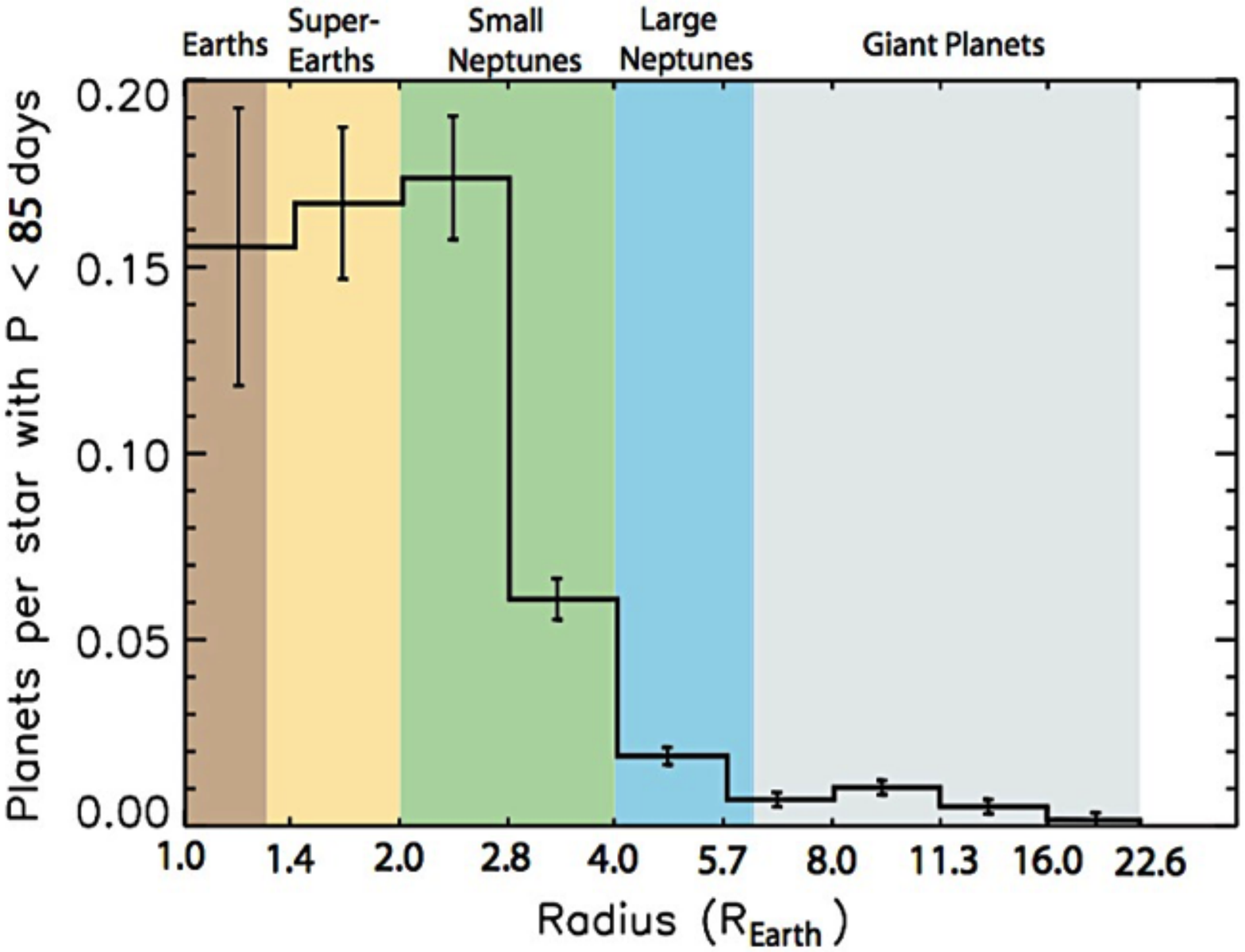}
\caption{Average number of planets per star and per size bin with an orbital period shorter than 85 days orbiting around F, G, K stars. The statistics was  extracted from the Q1 - Q6 Kepler data \citep{2013ApJ...766...81F}.\label{fig:FresFreq}}
\end{figure}

\noindent
The number of planets can be estimated as:
\begin{equation}
N_p = \frac{4}{3} \pi d^3 \rho_{*} P_{t, p} P_{geom}
\label{eq:num_plan}
\end{equation}
\noindent
where $d$ is the radius of a sphere  with the Sun at the centre, $\rho_{*}$ is the number density of the stars, $P_{t, p}$ is the probability of having a t-type planet orbiting with an orbital period $p$ (See Fig \ref{fig:FresFreq}). $P_{geom} =  R_{*} / a$ is the geometrical probability of a transit.

We simulated all the transiting planets in the solar system neighbourhood up to $m_K = 14$, all these planets described by $N_p$ constitute the ``Mission Reference Population''.

To include in the population sample the exoplanets known today, every time we predict a system with the same physical properties of a known system we replace it with the known one. In Sec \ref{sec:global_sample} we show that in the solar system neighbourhood there are $\sim$ 9500 planets for which the ARIEL science requirements can be reached in less that 6 transits or eclipses.

The equilibrium temperature (Eq \ref{eq:temp}) of the planet can be evaluated assuming  the incoming and outgoing radiation at the planetary surface are in equilibrium:

\begin{equation}
T_p = T_{*} \left( \frac{R_{*}}{2a} \right)^{\frac{1}{2}} \left( \frac{1-A}{\varepsilon} \right)^{\frac{1}{4}}
\label{eq:temp}
\end{equation}

Here $T_{*}$ and $R_{*}$ are the stellar temperature and radius,  $a$ the semi-major axis of the orbit, $A$ is the planetary albedo  and $\varepsilon$ is the atmospheric emissivity. 

The ARIEL space mission will focus on planets with an orbital period shorter than 50 days. As expected, shorter periods mean shorter semi-major axis and, therefore, from Eq \ref{eq:temp}, typically higher effective temperature.

\subsection{Planetary masses and densities}
To simulate a realistic planetary population we need to consider a distribution of masses given a planetary radius. The planetary mass affects directly the surface gravity and therefore the scale height ($H$) of the atmosphere:

\begin{equation}
H = \frac{k \, T}{\mu \, g}
\label{eq:high_scale}
\end{equation}

\begin{figure}[!htbp]
\centering
\includegraphics[scale=0.4, angle=-90]{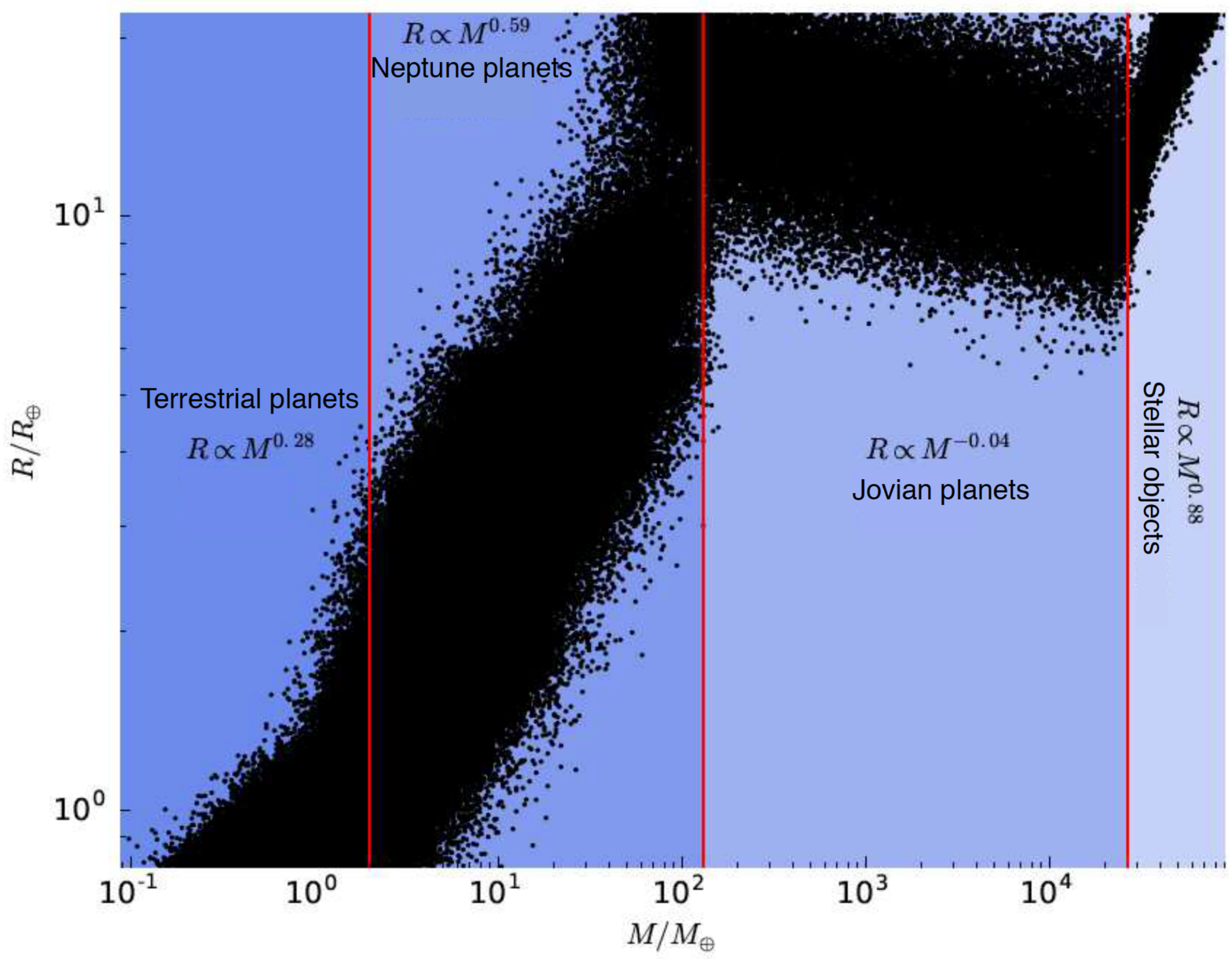}
\caption{Mass-Radius distribution for all the simulated planets. The mass-radius relationship has been calculated with the \citet{2016arXiv160308614C} tool.}
\label{fig:mass_chen}
\end{figure}

The mass estimate is not a trivial task, given the range of planetary densities observed today. We used a Python tool written by \citet{2016arXiv160308614C} to estimate the mass of all the planets in our simulated sample. In the ARIEL planetary sample there are both known and simulated planets. \citet{2016arXiv160308614C} use the currently known planets to derive the statistical distribution of the mass of a given planet when its radius is known. Thus, for each planet in our initial sample, the mass is randomly drawn following this distribution except for known systems. In Fig \ref{fig:mass_chen} we show the mass distribution for all the planets in our simulations. Moreover, as a very few planets have a radius larger than $20R_{\oplus}$, we use that radius as an upper limit. There is already a well known degeneracy in the $7-20R_{\oplus}$ range: objects with a radius within that range can be planets as well as very cool stars. However, this should not be too concerning, as observations have shown that very short-period, low-mass stellar companions are much less frequent than hot giant planets \citep{2015ApJ...814..148P}.

\clearpage

\section{ARIEL science goals and Mission Reference Population}\label{sec:global_sample}
\subsection{The 3 tier approach}
The ARIEL primary science objectives call for atmospheric spectra or photometric lightcurves of a large and diverse sample of known exoplanets covering a wide range of masses, densities, equilibrium temperatures, orbital properties and host-stars. Other science objectives require, by contrast, the very deep knowledge of a select sub-sample of objects. To maximise the science return of ARIEL and take full advantage of its unique characteristics, a three-tiered approach has been considered, where three different samples are observed at optimised spectral resolutions, wavelength intervals and signal-to-noise ratios. (a summary of the three-tiers and observational methods is given below in table \ref{tab:ARIEL_3-tiers}). 

In this section we present the pool of potential targets that could reach the specifications for each tier in a reasonable number of observations. The number of targets for the various Tiers are shown as a function of planetary radius in Fig  \ref{fig:Survey_all planets}, \ref{fig:Deep_all planets} and \ref{fig:Benchmark_all planets} and as a function of effective temperature in  \ref{fig:Survey_all planets_temperature}, \ref{fig:Deep_all planets_temperature} and \ref{fig:Benchmark_all planets_temperature}. Note that the planets shown in these figures do not represent the final sample, as it would take too long to observe all of them. They are the pool from which the MRS can be selected to best address the scientific questions summarized below. The fact that the number of potential targets is much larger than the number that can be observed illustrates that ARIEL can choose the final sample among a great variety of observable planets, providing a lot a flexibility.

\begin{table}[!htbp]
\centering
\begin{tabular}{|p{3cm}|p{6cm}|}
  \hline
  \multicolumn{2}{|c|}{ARIEL 3-tiers} \\
  \hline
  Survey ($\sim$37\%) & Low Spectral Resolution observations of a large sample of planets in the Vis-IR, with SNR $\ge$7  \\
\hline
  Deep ($\sim$60\%) & Higher Spectral Resolution observations of a sub-sample in the VIS-IR \\
\hline
  \footnotesize{Benchmark ($\sim$3\%)} & Very best planets, re-observed multiple time with all techniques \\
\hline
\end{tabular}
\caption{Summary of the survey tiers and the observational methods they will be accomplished. Each tier is expressed in terms of nominal mission lifetime ARIEL could spend on them.\label{tab:ARIEL_3-tiers}}
\end{table}

The key questions and objectives of each tier can be summarised as follows (see Tinetti et al., in prep. for further details):

\textbf{Survey}:

\begin{itemize}
\item \emph{What fraction of planets are covered by clouds? } -- Tier 1 mode is particularly useful for discriminating between planets that are likely to have clear atmospheres, versus those that are so cloudy that no molecular absorption features are visible in transmission. Extremely cloudy planets may be identified simply from low-resolution observations over a broad wavelength range. This preliminary information will therefore allow us to take an informed decision about whether to continue the spectral characterization of the planet at higher spectral resolution, and therefore include or not the planet in the Tier 2 sample.
\item \emph{What fraction of small planets have still hydrogen and helium retained from the protoplanetary disk?} -- Primordial (primary atmosphere) atmospheres are expected to be mainly made of hydrogen and helium, i.e. the gaseous composition of the protoplanetary nebula. If an atmosphere is made of heavier elements, then the atmosphere has probably evolved (secondary atmosphere). An easy way to distinguish between primordial (hydrogen-rich) and evolved atmospheres (metal-rich), is to examine the transit spectra of the planet: the main atmospheric component will influence the atmospheric scale height, thus changing noticeably the amplitude of the spectral features. This question is essential to understand how super-Earths formed and evolved.
\item \emph{Can we classify planets through colour-colour diagrams or colour-magnitude diagrams?} --
Colour-colour or colour-magnitude diagrams are a traditional way of comparing and categorising luminous objects in astronomy. Similarly to the Herzsprung-Russell diagram, which led to a breakthrough in understanding stellar formation and evolution, the compilation of similar diagrams for exoplanets might lead to similar developments \citep{2014MNRAS.444..711T}.
\item  \emph{What is the bulk composition of the terrestrial exoplanets?} -- The planetary density may constrain the composition of the planet interior. However this measurement alone may lead to non-unique interpretations \citep{2007ApJ...665.1413V}. A robust determination of the composition of the upper atmosphere of transiting planets will reveal the extent of compositional segregation between the atmosphere and the interior, removing the degeneracy originating from the uncertainty in the presence and mass of their (inflated?) atmospheres. 
\item \emph{What is the energy balance of the planet?} --
Eclipse photometric measurements in the optical and infrared  can provide the bulk temperature and albedo of the planet, thereby allowing the estimation of the planetary energy balance and whether the planet has an internal heat source or not.
\end{itemize}

\textbf{Deep}:

A key objective of ARIEL is to understand whether there is a correlation between the chemistry of the planet and basic parameters such as planetary size, density, temperature and stellar type and metallicity. Spectroscopic measurements at higher resolution will allow in particular to measure:
\begin{itemize}
\item The main atmospheric component for small planets;
\item The chemical abundances of trace gases, which is pivotal to understand the type of chemistry (equilibrum/non equilibrium). 
\item The atmospheric thermal structure, both vertical and horizontal;
\item The cloud properties, i.e. cloud particles size and distribution, 
\item The elemental composition in gaseous planets. This information can be used to constrain formation scenarios \citep{2011ApJ...740..109O}.
\end{itemize}

\textbf{Benchmark}:

A fraction of  planets around very bright stars will be observed repeatedly through time to obtain:
\begin{itemize}
\item A very detailed knowledge of the planetary chemistry and dynamics;
\item An understanding of the weather, and the spatial and temporal variability of the atmosphere.
\end{itemize} 
Benchmark planets are the best candidates for phase-curve spectroscopic measurements.

\subsection{Key science questions}\label{sec:survey_potential}
In this section we show a full list of potential targets for ARIEL: these are expected to evolve  until launch, and will be updated regularly to include new  planet discoveries.

\begin{figure}[!htbp]
\centering
\includegraphics[scale=0.22]{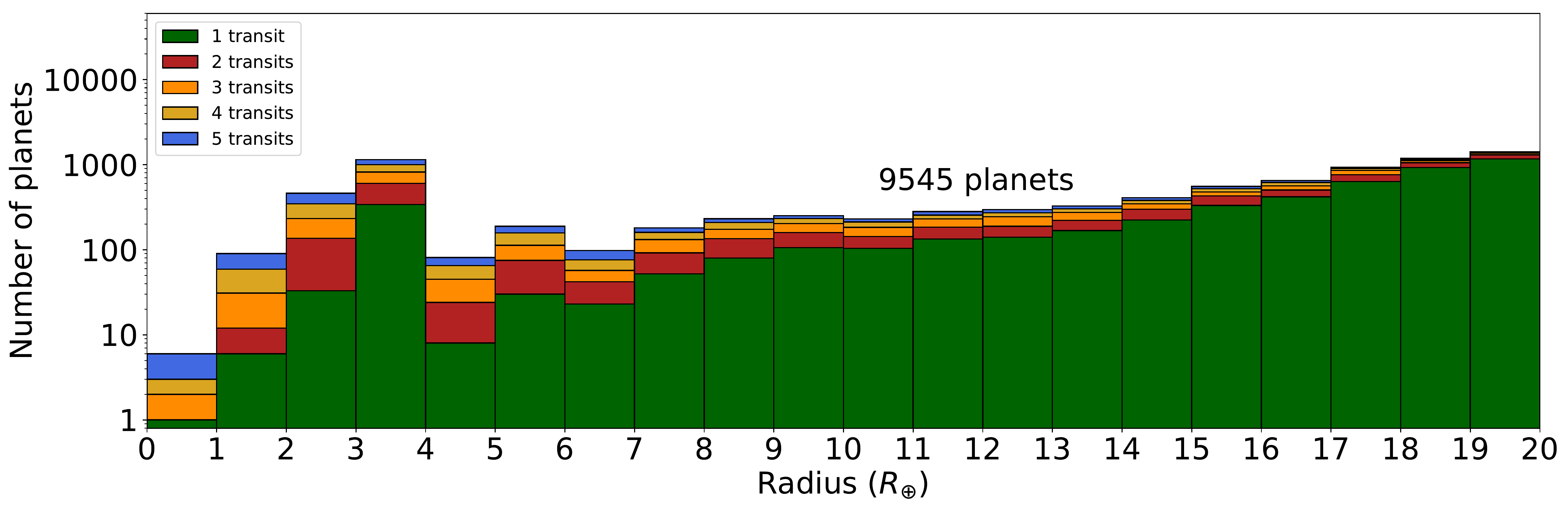}
\caption{Complete set of Tier 1 planets from the ARIEL missione reference population. The final list of Tier 1 planets will include an optimal sub-sample. Different colours indicate the number of transits/eclipses needed to reach Tier 1 performances. The planets shown here can achieve the Tier 1 requirements combining the signal of $\le 5$ transits/eclipses.}
\label{fig:Survey_all planets}
\end{figure}

\begin{figure}[!htbp]
\centering
\includegraphics[scale=0.22]{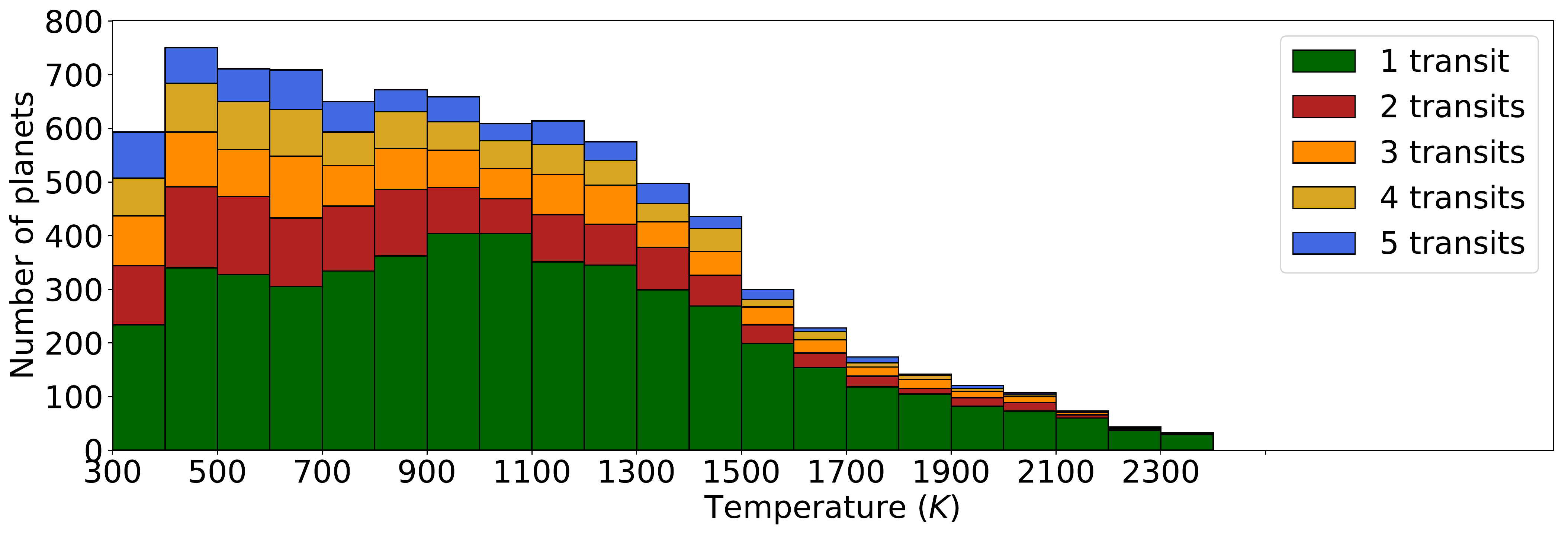}
\caption{Temperature distribution for the planets illustrated in fig. \ref{fig:Survey_all planets}. }
\label{fig:Survey_all planets_temperature}
\end{figure}

ARIEL Tier 1 (Survey) will analyse a large sample of exoplanets to address science questions where a statistically significant population of objects needs to be observed. ARIEL Tier 1 will also allow a rapid, broad characterisation of planets permitting a more informed selection of Tier 2 and Tier 3 planetary candidates. For most Tier 1 planetary candidates, Tier 1 performances can be reached between 1 and 2 transits/eclipses. In Fig \ref{fig:Survey_all planets} and \ref{fig:Survey_all planets_temperature} we show that in the solar system neighbourhood there are $\sim$ 9500 observable by ARIEL for which the science requirements can be reached in less than 6 transits or eclipses.

ARIEL Tier 2 (Deep, the core of the mission) will analyse a sub-sample of Tier 1 planets with a higher spectral resolution, allowing an optimal characterisation of the atmospheres, including information on the thermal structure, abundance of trace gases, clouds and elemental composition. 

\begin{figure}[!htbp]
\centering
\includegraphics[scale=0.22]{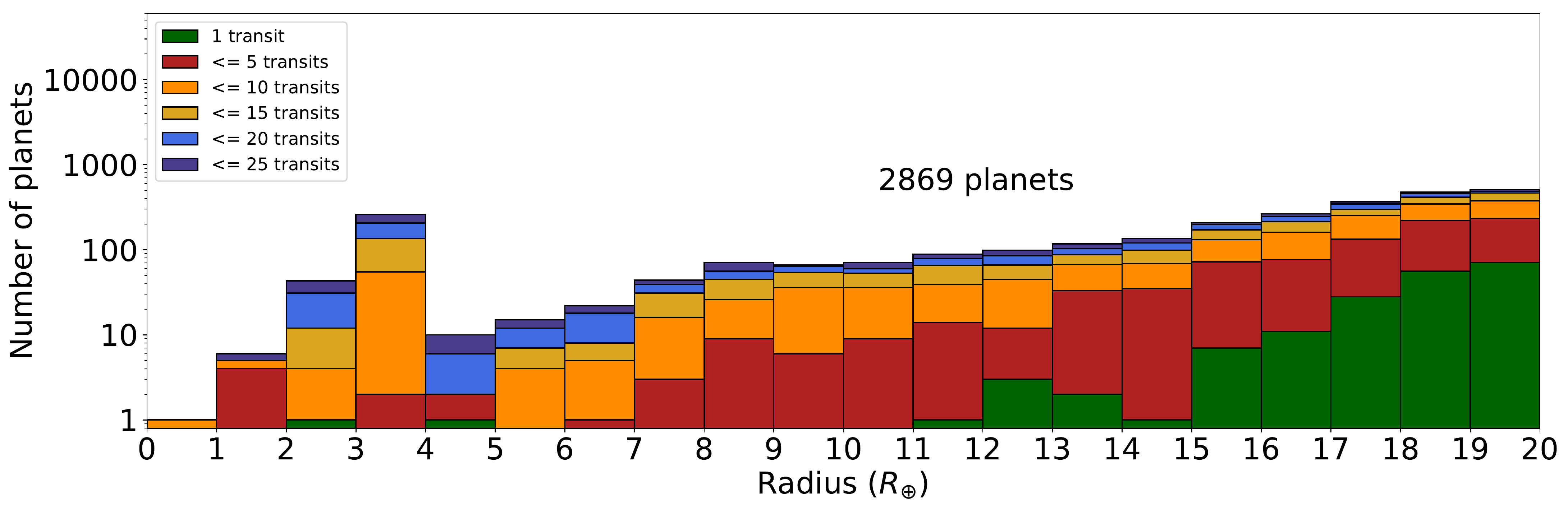}
\caption{Planets from the ARIEL mission reference population in the Deep mode (Tier 2) with a small/moderate number of transits/eclipses, divided in size bins. The final list of Tier 2 planets will include an optimal sub-sample. Different colours indicate the number of transits/eclipses needed to reach Tier 2 performances. }
\label{fig:Deep_all planets}
\end{figure}

\begin{figure}[!htbp]
\centering
\includegraphics[scale=0.22]{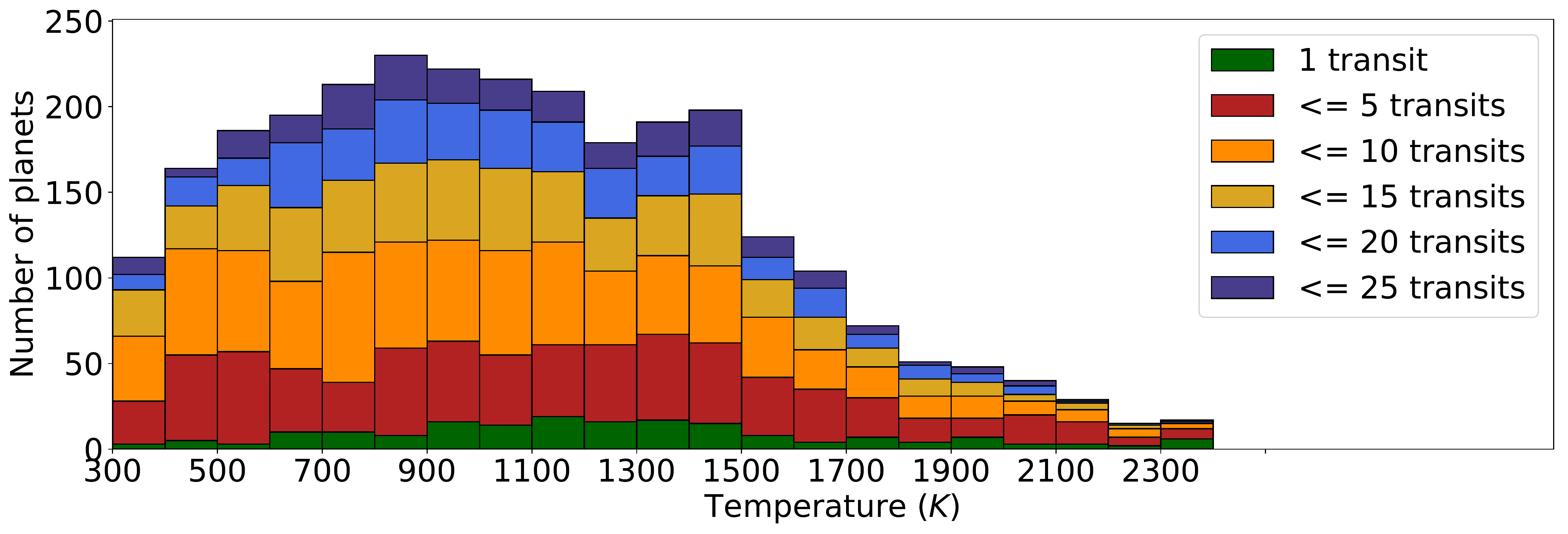}
\caption{Temperature distribution for the planets illustrated in fig.  \ref{fig:Deep_all planets}.}
\label{fig:Deep_all planets_temperature}
\end{figure}

In Fig \ref{fig:Deep_all planets} and \ref{fig:Deep_all planets_temperature} we show the properties of all the planetary candidates that could be studied by ARIEL in the Deep mode with a small/moderate number of transit or eclipse events. 

\begin{figure}[!htbp]
\centering
\includegraphics[scale=0.22]{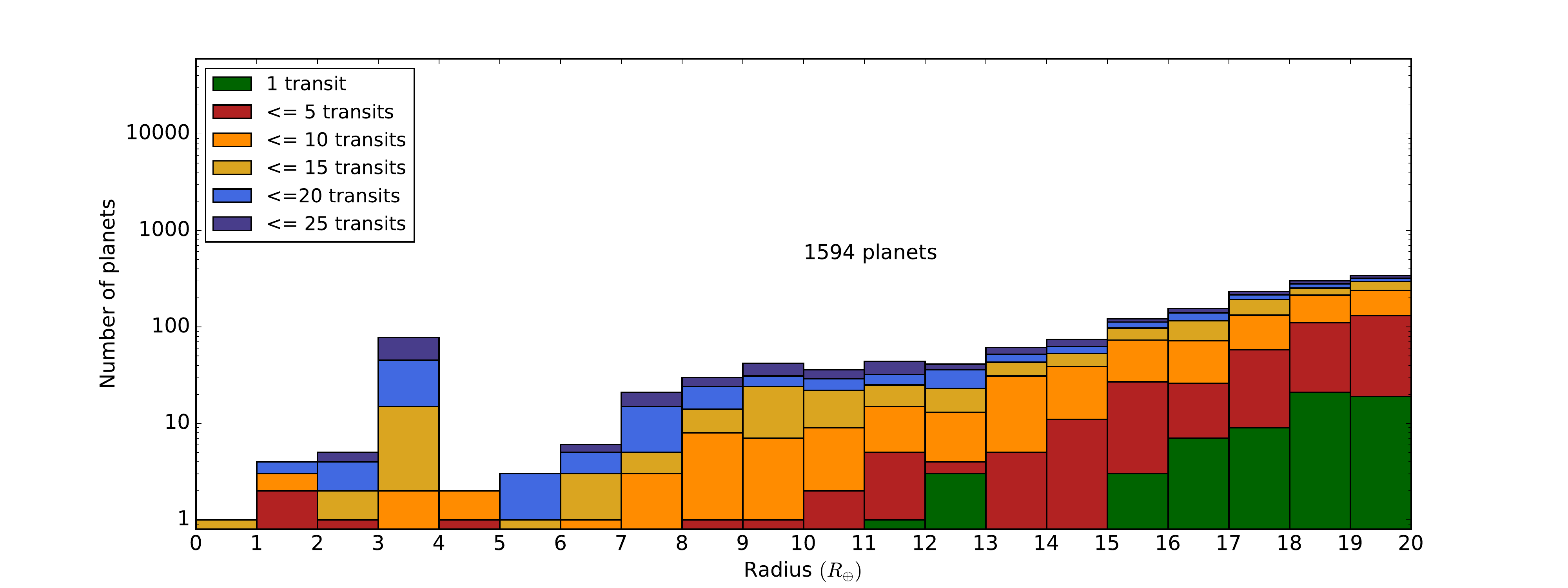}
\caption{Number of planets from the mission reference population observable by ARIEL in the Benchmark mode with a $< 25$ number of transits/eclipses, divided in size bins. Different colours indicate the number of transits/eclipses needed to reach Tier 3 performances.}
\label{fig:Benchmark_all planets}
\end{figure}

\begin{figure}[!htbp]
\centering
\includegraphics[scale=0.22]{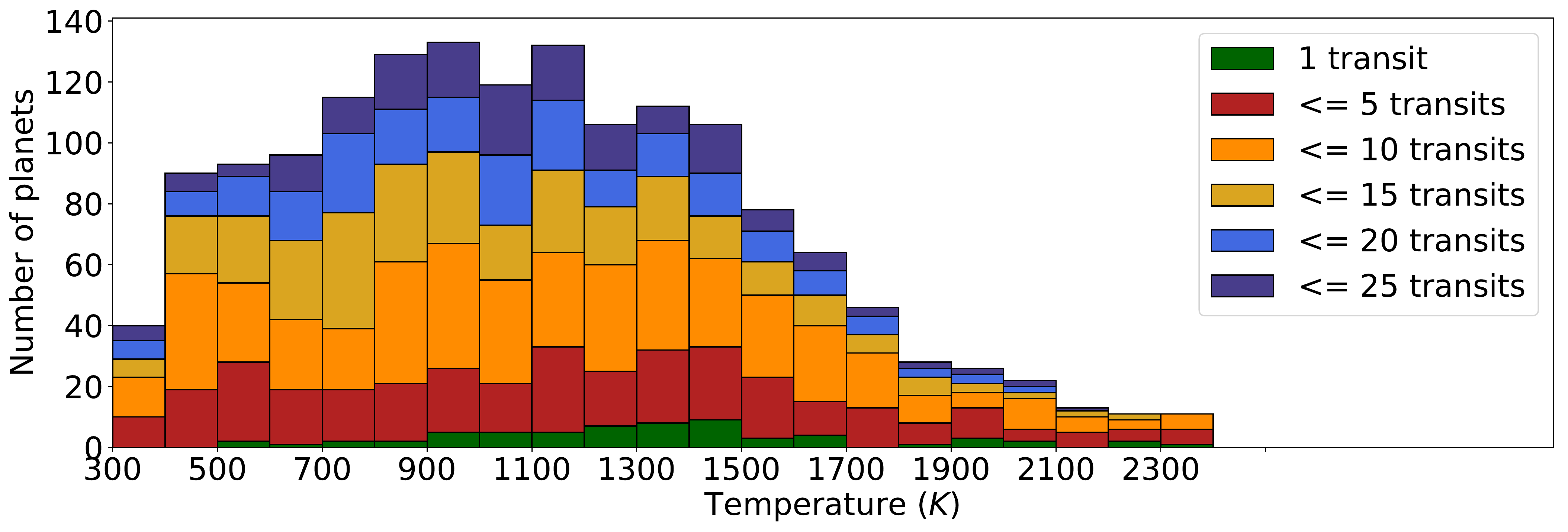}
\caption{Temperature distribution for the planets illustrated in fig.  \ref{fig:Benchmark_all planets}. }
\label{fig:Benchmark_all planets_temperature}
\end{figure}

The third ARIEL Tier (Benchmark, the reference planets) will study the best planets (section \ref{sec:benchmark}), i.e. the ones orbiting very bright stars which can be studied in full spectral resolution with a relatively small number of transits/eclipses.  For the planets observed in benchmark mode in 1 or 2 events, it is possible to study the spatial and temporal variability (i.e. study the weather and evaluate its impact when observations are averaged over time). In Fig \ref{fig:Benchmark_all planets} and \ref{fig:Benchmark_all planets_temperature} we show the properties of the Tier 3 planetary candidates.

\clearpage

\section{A possible scenario for the ARIEL space mission}\label{sec:possible_MRS}
In Section \ref{sec:global_sample} we presented a comprehensive list of planet candidates which could be observed with the ARIEL space mission. 
Here we discuss possible optimisations of the Mission Reference Sample, which ideally should include a large and diverse sample of planets, have the right balance among the three Tiers and, most importantly, must be completed during the nominal mission lifetime (4 years including the commissioning phase).

\begin{figure}[!htbp]
\centering
\includegraphics[scale=0.3]{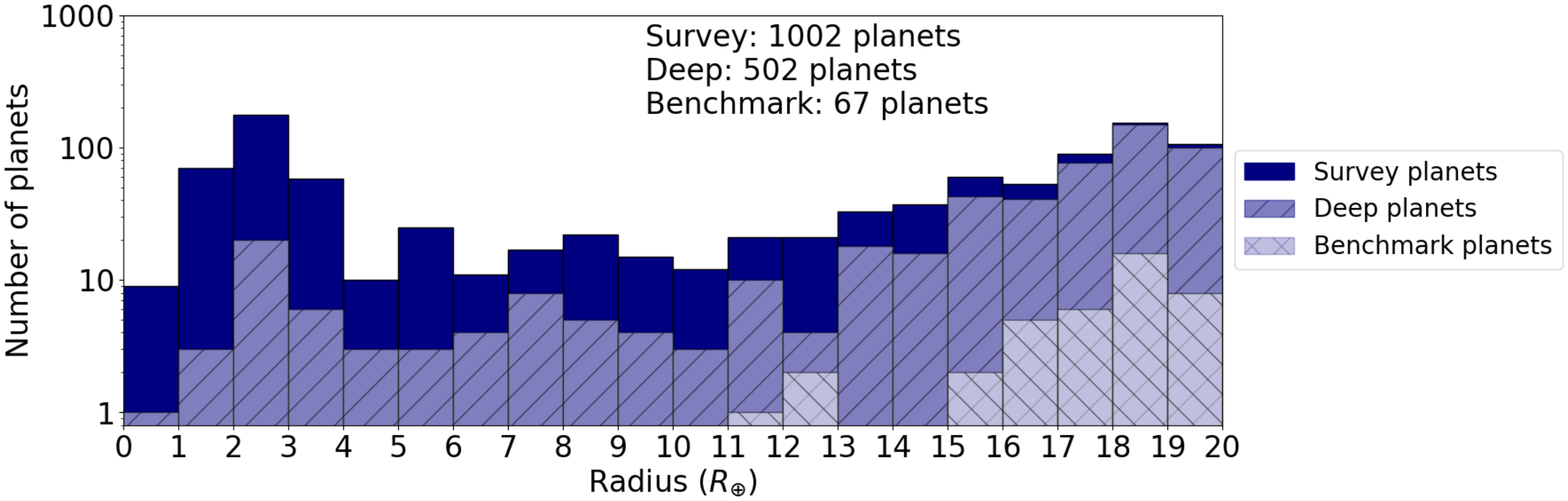}
\caption{Overview of the ARIEL MRS, comparing the number of planets observable in the three tiers during the mission lifetime.  }
\label{fig:tier_superimposition}
\end{figure}
In Fig \ref{fig:tier_superimposition} we show a possible MRS with all the three tiers nested  together. The first MRS shows the maximum number of targets, taking into account the nominal mission lifetime. It has been built starting from all the targets feasible within one transit/eclipse, and adding all the targets that can be done within 2, 3, 4 and so on transits/eclipses in ascending order. This is just one of the possible configurations for the MRS, and one would expect the ARIEL MRS to evolve in response of new exoplanetary discoveries in the next decade.

\subsection{MRS Tier 1: Survey}\label{sec:survey}
\begin{figure}[!htbp]
\centering
\includegraphics[scale=0.22]{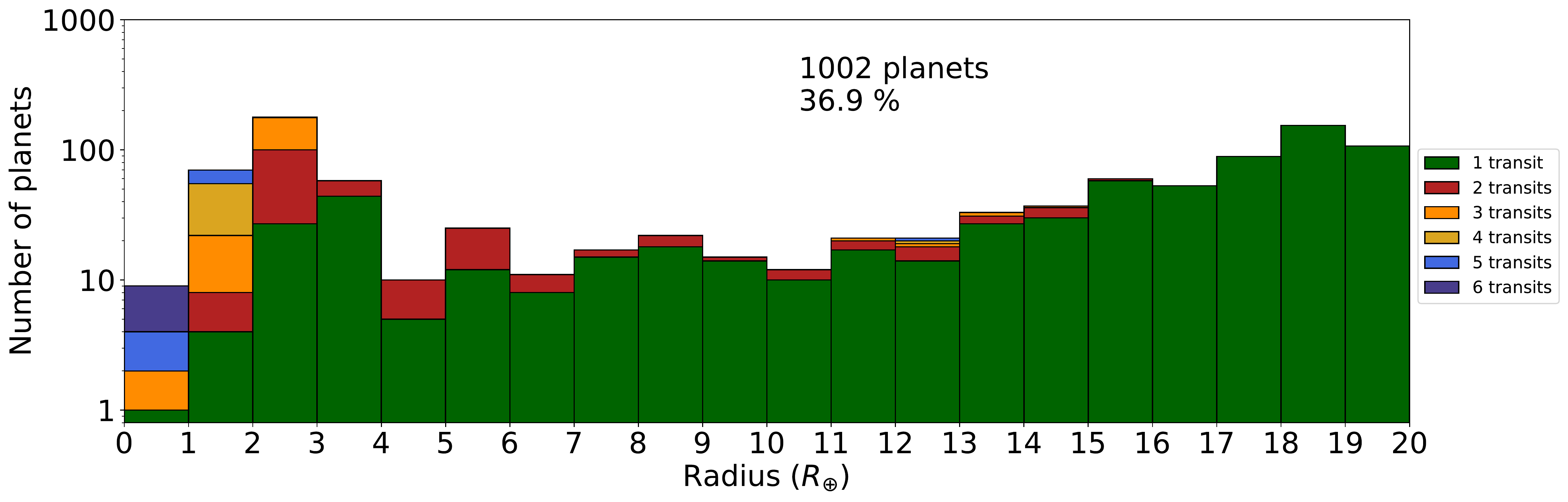}
\caption{ARIEL MRS Tier 1 planets organised in size-bins. Different colours indicate the number of transits/eclipses needed to reach Tier 1 performances.
}
\label{fig:Survey}
\end{figure}

\begin{figure}[!htbp]
\centering
\includegraphics[scale=0.22]{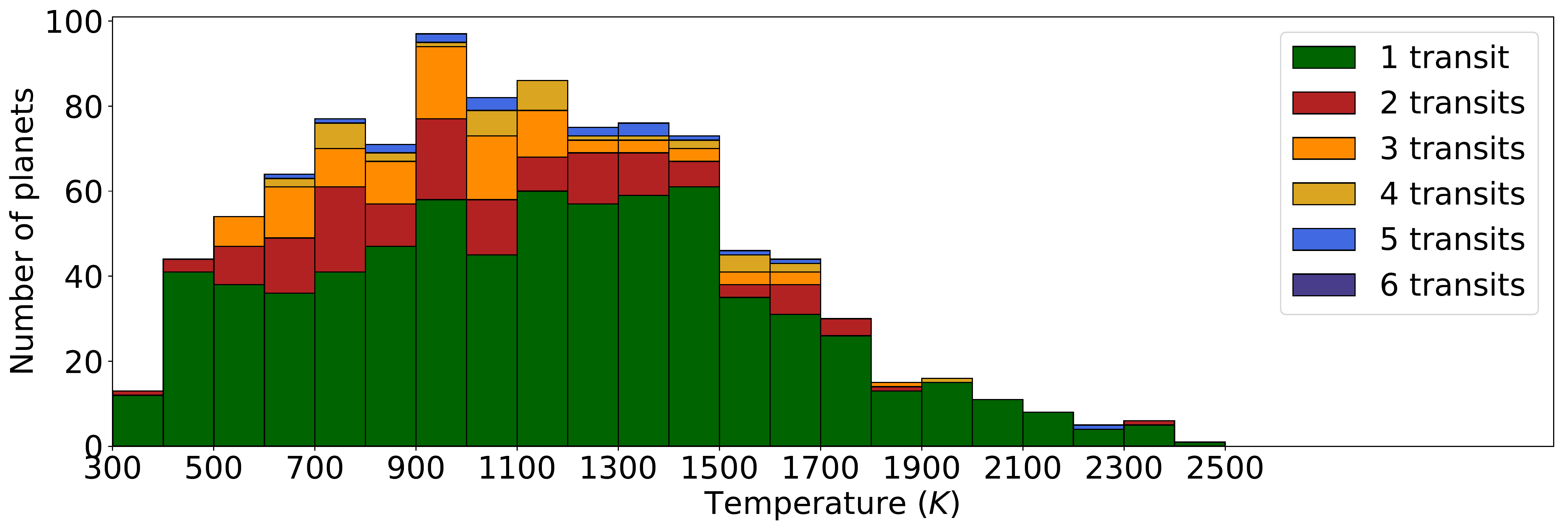}
\caption{ARIEL MRS Tier 1 planets organised in temperature-bins. Different colours indicate the number of transits/eclipses needed to reach Tier 1 performances.}
\label{fig:Survey_temperature}
\end{figure}

\begin{figure}[!htbp]
\centering
\includegraphics[scale=0.22]{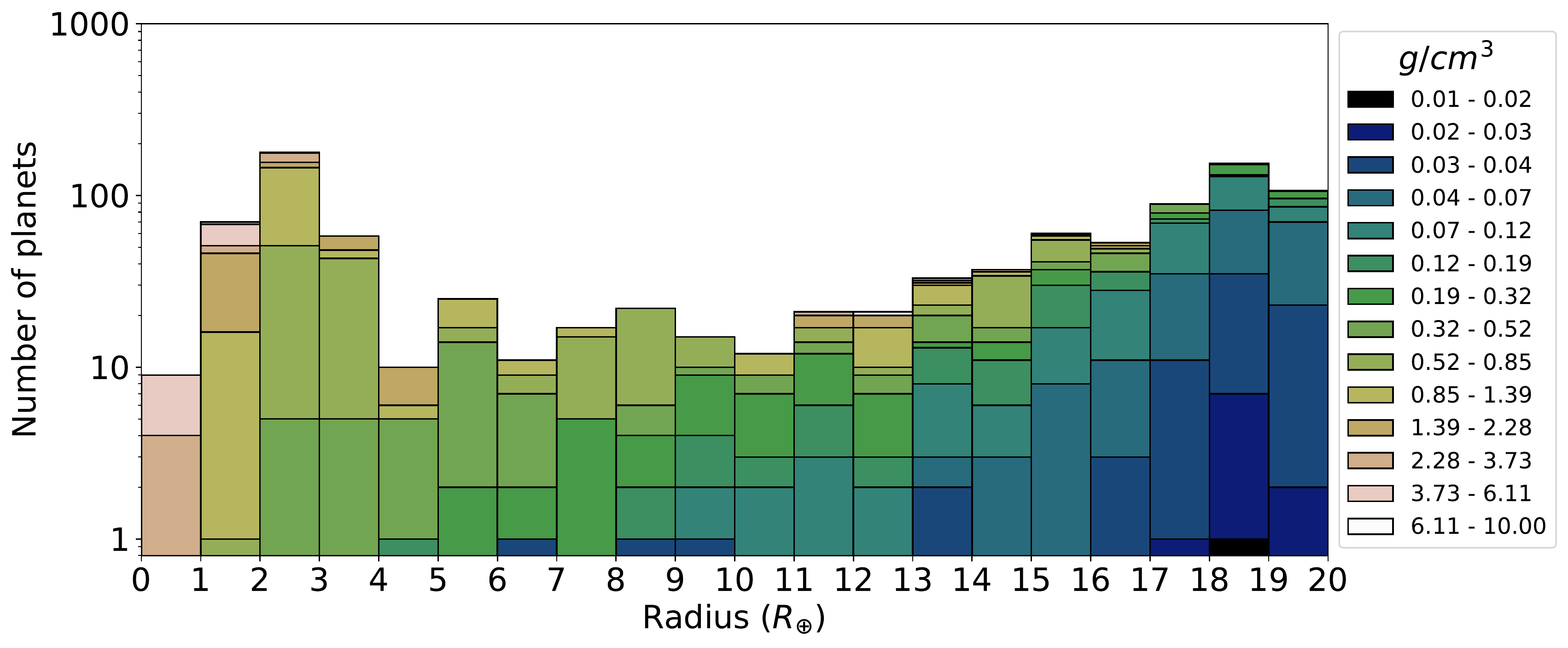}
\caption{ARIEL MRS Tier 1 planets organised in size-bins. Different colours indicate differences in the simulated planetary densities.}
\label{fig:Survey_densities}
\end{figure}

\begin{figure}[!htbp]
\centering
\includegraphics[scale=0.22]{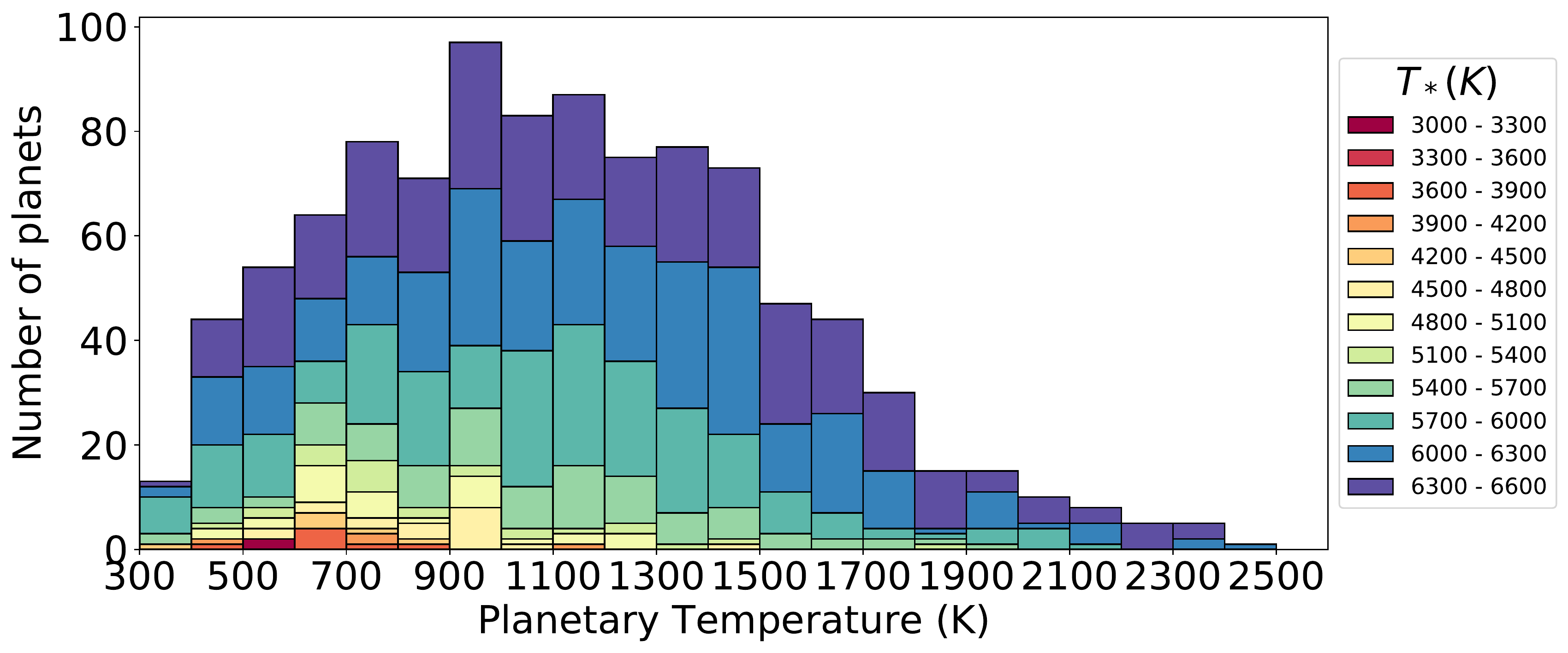}
\caption{ARIEL MRS Tier 1 planets organised in temperature-bins. Different colours indicate differences in the simulated stellar temperatures.}
\label{fig:Survey_star_temperature}
\end{figure}

Our simulations indicate that the current ARIEL design as presented at the end of the Phase A study, allows to observe 1002 planets in Tier 1. All the planets can be observed in 1538 satellite visits i.e. 37\% of the mission lifetime. Most giant planets and Neptunes fulfil the Tier 1 science objectives in 1 transit/eclipse, the smaller planets require up to 6 events (fig. \ref{fig:Survey} and \ref{fig:Survey_temperature} ). Fig. \ref{fig:Survey_densities} and \ref{fig:Survey_star_temperature} illustrate how the 1002 planets are distributed in terms of planetary size, temperature, density and stellar type.

\subsection{MRS Tier 2: Deep}\label{sec:deep}
\begin{figure}[!htbp]
\centering
\includegraphics[scale=0.22]{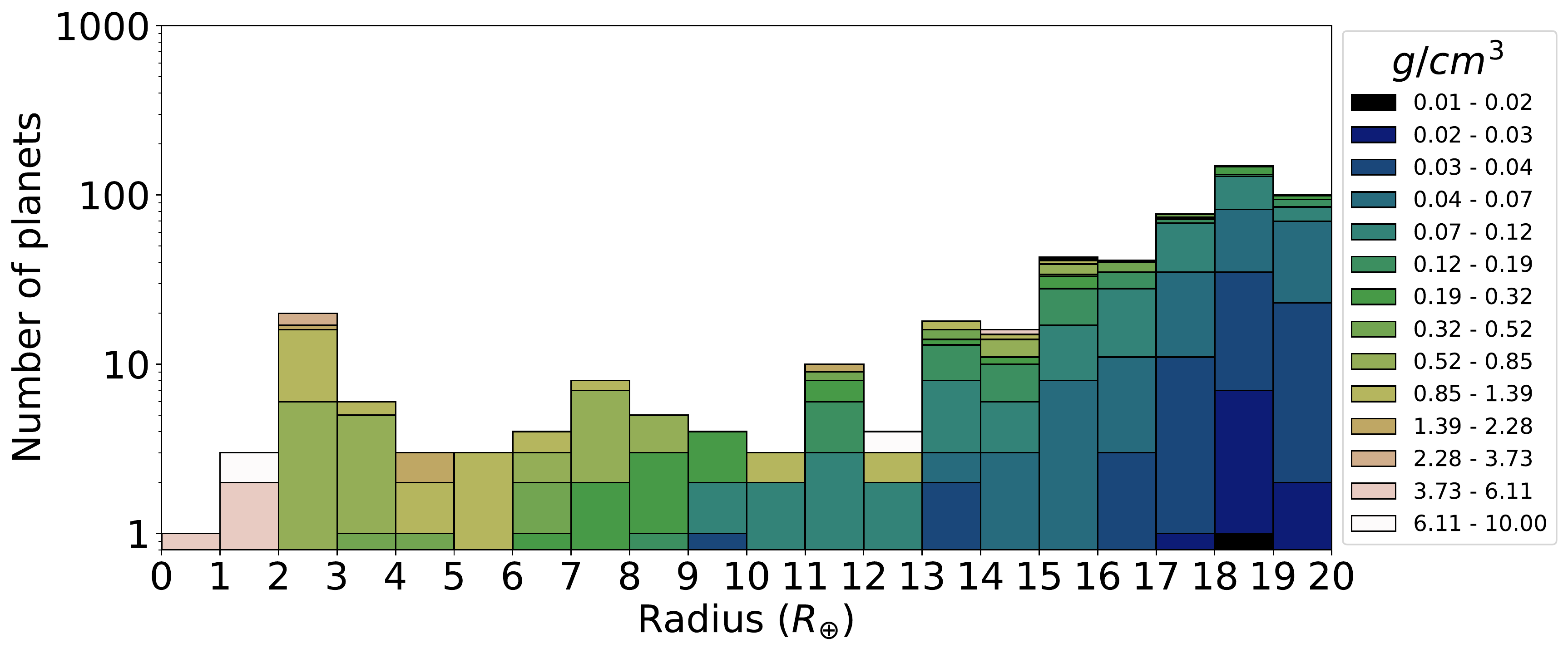}
\caption{ARIEL MRS Tier 2 planets organised in size-bins. Different colours indicate differences in the simulated planetary densities.}
\label{fig:Deep_densities}
\end{figure}

\begin{figure}[!htbp]
\centering
\includegraphics[scale=0.22]{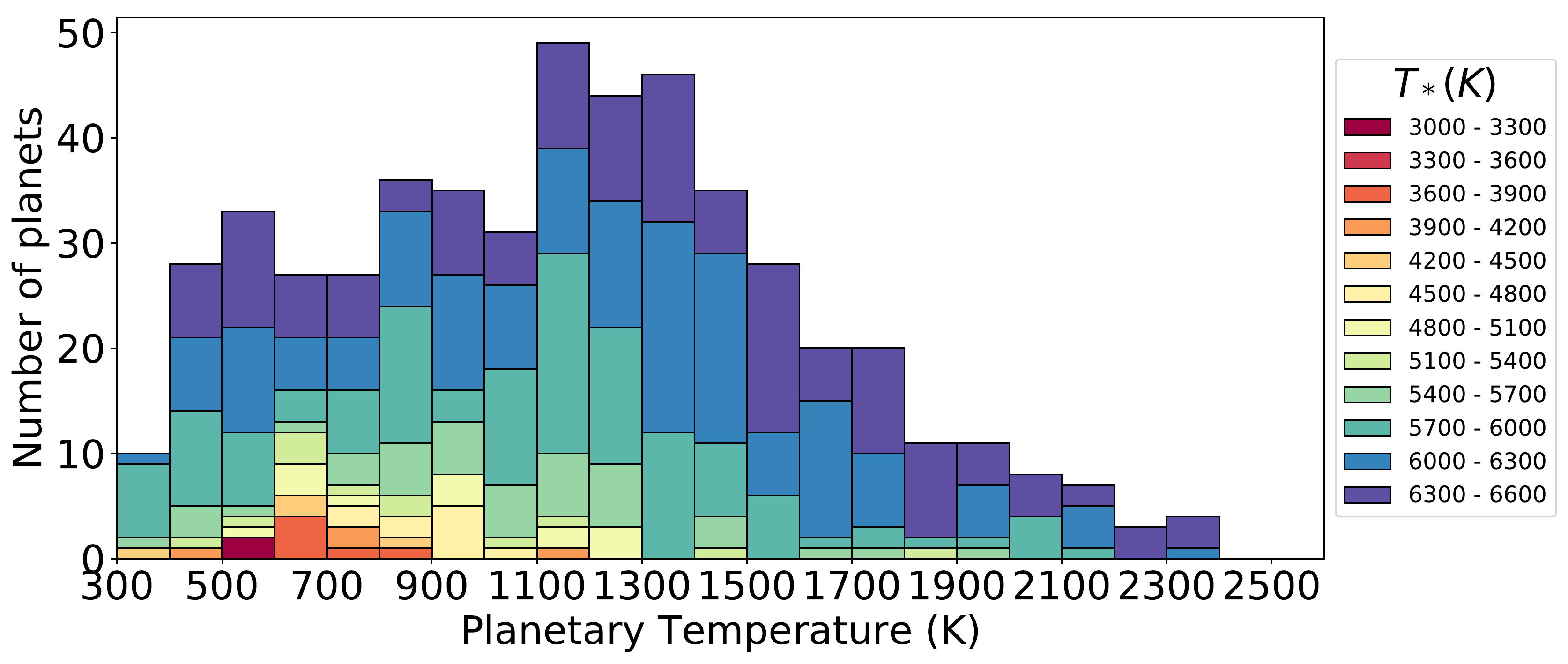}
\caption{ARIEL MRS Tier 2 planets organised in temperature-bins. Different colours indicate differences in the simulated stellar temperatures.}
\label{fig:Deep_star_temperature}
\end{figure}

The Deep is the core of the mission. Our simulations indicate that the current ARIEL design as presented at the end of the Phase A study, allows to observe $\sim$ 500 planets in Tier 2 assuming 60\% of the mission lifetime. Fig. \ref{fig:Deep_densities} and \ref{fig:Deep_star_temperature} illustrate how the 500 planets are distributed in terms of planetary size, temperature, density and stellar type. 

\begin{figure}[!htbp]
\centering
\includegraphics[scale=0.22]{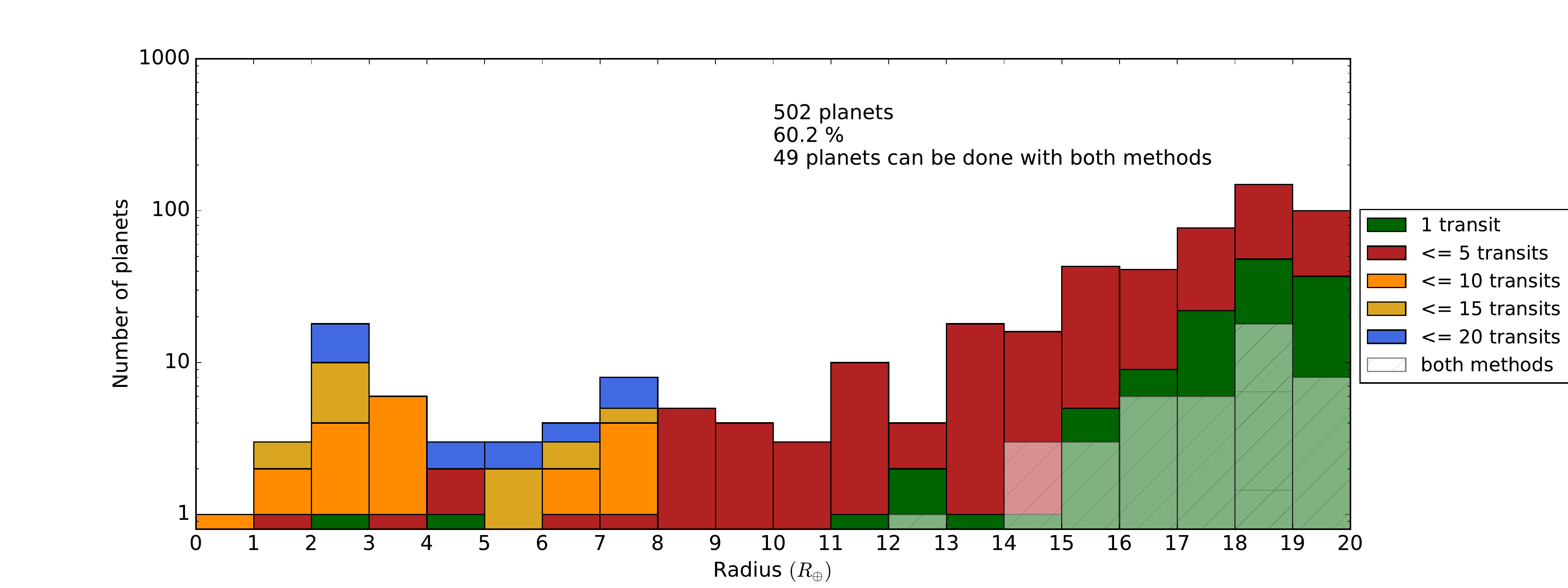}
\caption{ARIEL MRS Tier 2 planets organised in size-bins. Different colours indicate the number of transits/eclipses needed to reach Tier 2 performances.
Stripes indicate planets that will be studied with both transit and eclipse methods}
\label{fig:Deep}
\end{figure}

\begin{figure}[!htbp]
\centering
\includegraphics[scale=0.22]{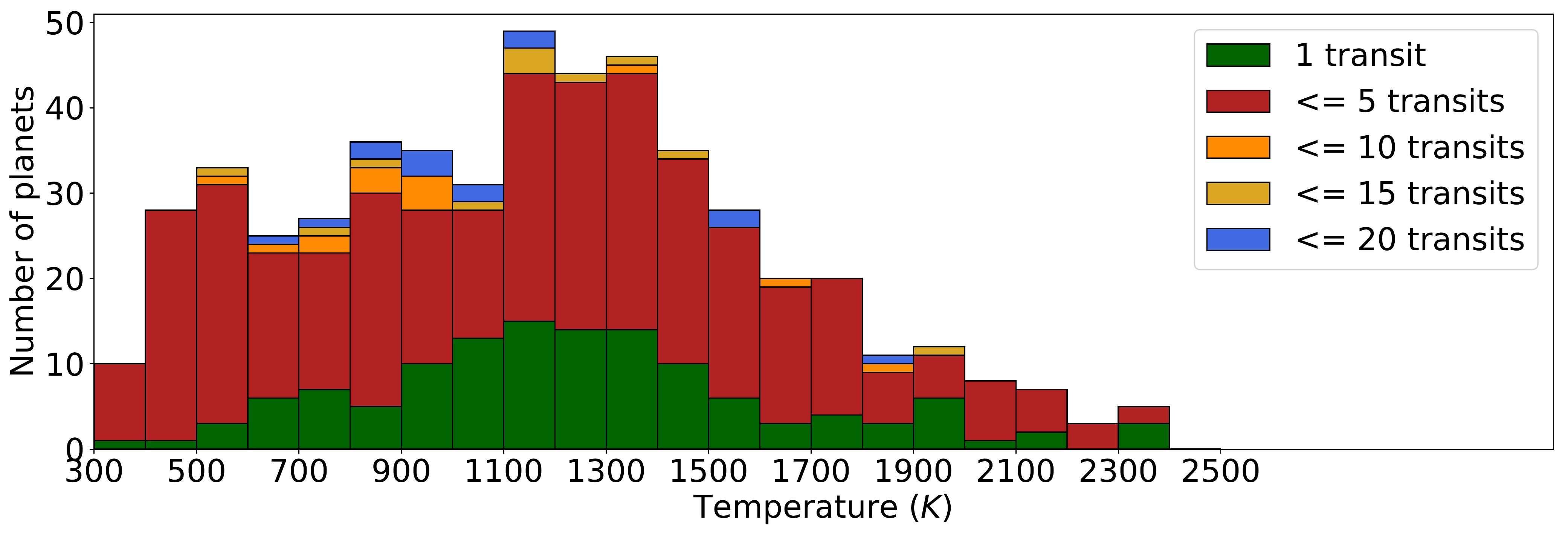}
\caption{ARIEL MRS Tier 2 planets organised in temperature-bins. Different colours indicate the number of transits/eclipses needed to reach Tier 2 performances.}
\label{fig:Deep_temperature}
\end{figure}

Most gaseous planets fulfil the Tier 2 science objectives in less than five transits/eclipses, the small planets require up to twenty events (fig. \ref{fig:Deep} and \ref{fig:Deep_temperature} ).
We included a variety of planets from cold (300 K) to very hot (2500 K) as shown in Fig \ref{fig:Deep_temperature}.
We scheduled also $\sim$ 50 planets that will be studied with both transit and eclipse methods, indicated by stripes in Fig \ref{fig:Deep}). These are the best candidates for phase-curves observations, which can be included at the expenses of  the number of  Tier 2 planets observed.

\clearpage

\subsection{MRS Tier 3: Benchmark}\label{sec:benchmark}
\begin{figure}[!htbp]
\centering
\includegraphics[scale=0.22]{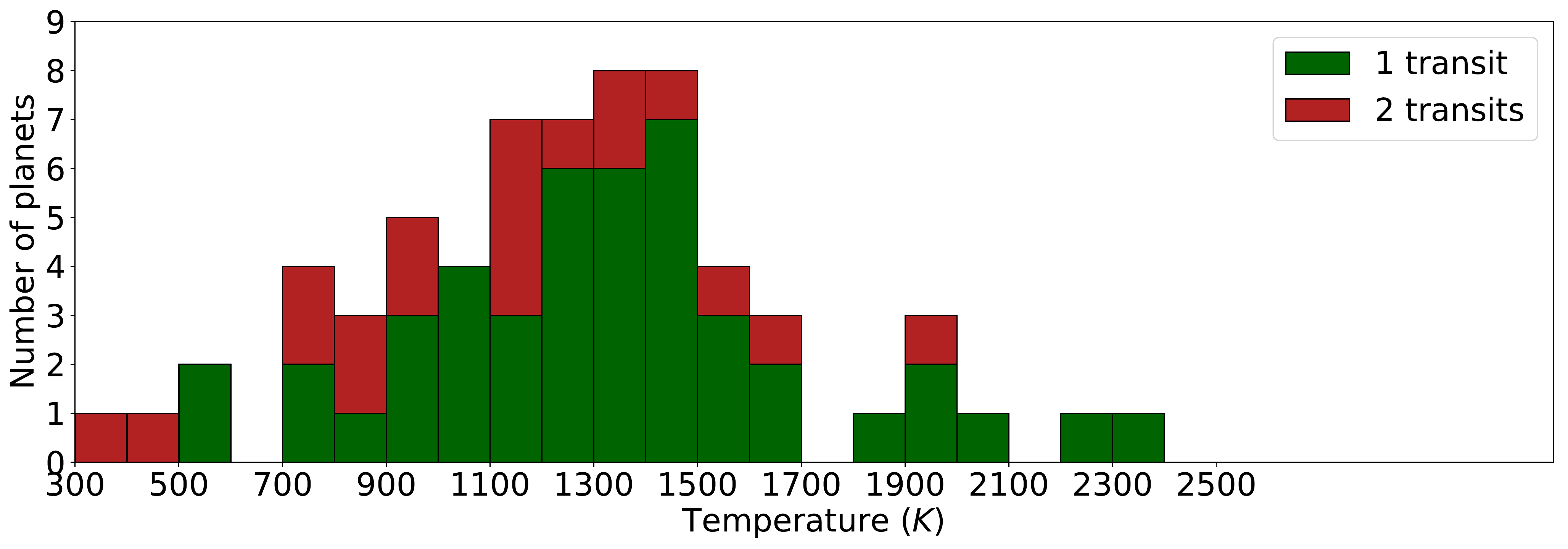}
\caption{Temperature distribution of the planets observable by ARIEL in the Benchmark.}
\label{fig:Benchmark_temperature}
\end{figure}

In the current MRS, we have selected as Tier 3, 67 gaseous planets for weather studies. Fig. (\ref{fig:Benchmark_temperature}) shows the temperature distribution covered by the Tier 3 sample. Only 3\% of the mission lifetime is required to achieve the Tier 3 science objectives for this sample.

\clearpage

\subsection{Compliance with TESS expected yields}\label{sec:TESS_mission}
\begin{figure}[!htbp]
\centering
\includegraphics[scale=0.22]{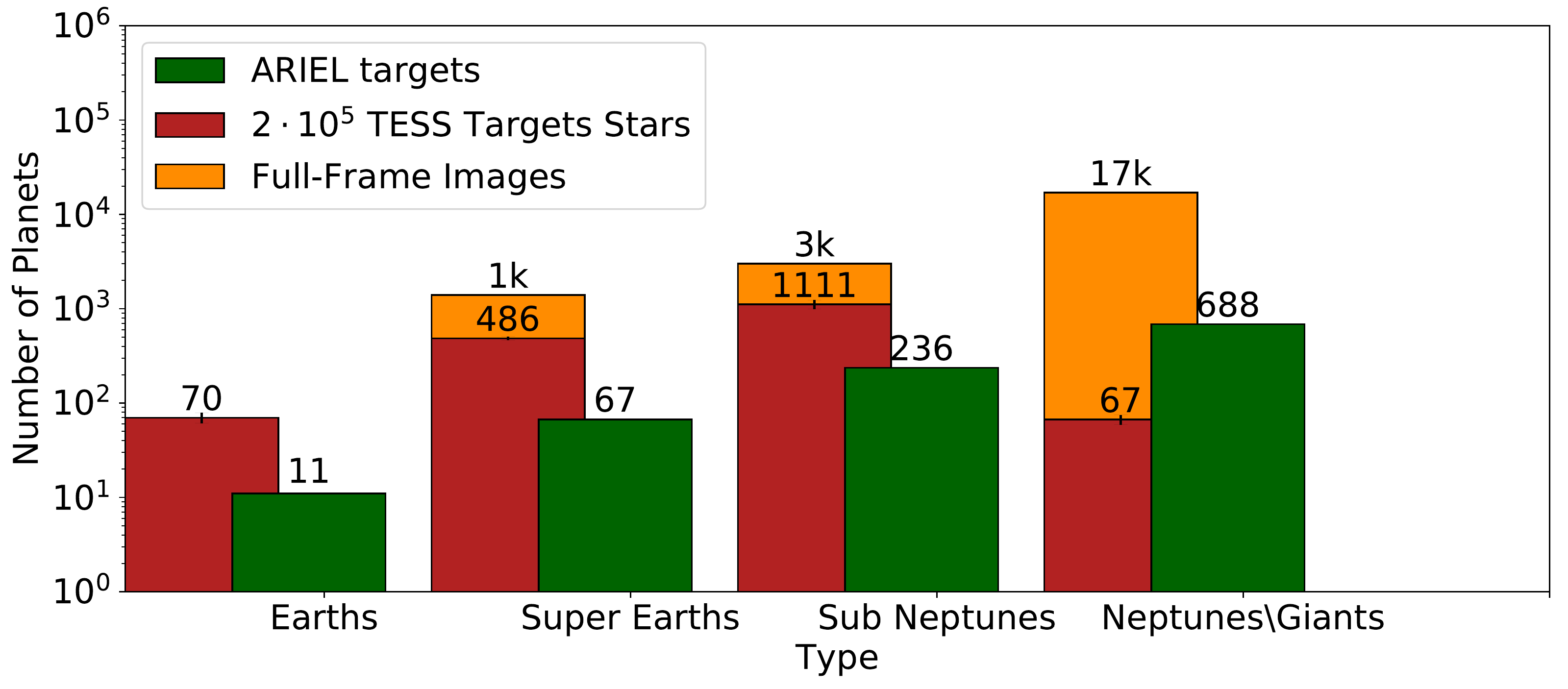}
\caption{Comparison between the TESS targets \citep{2015ApJ...809...77S} and the ARIEL MRS (green bars). }
\label{fig:TESS_comparison}
\end{figure}

The Transiting Exoplanet Survey Satellite (TESS) is expected to provide a large fraction of the targets observable by ARIEL. The numbers of targets envisioned in the sample presented here are perfectly in line with the expected yield from The Transiting Exoplanet Survey Satellite (TESS), as shown in Fig \ref{fig:TESS_comparison} where we compare the expected TESS discoveries and the ARIEL MRS. We see that the ARIEL MRS is well within the TESS sample \citep{2015ApJ...809...77S}. The success of the TESS mission will allow the characterisation of hundreds of planets by ARIEL.

\subsection{ARIEL MRS with currently known targets }\label{subsec:known_planets}
\begin{figure}[!htbp]
\centering
\includegraphics[scale=0.22]{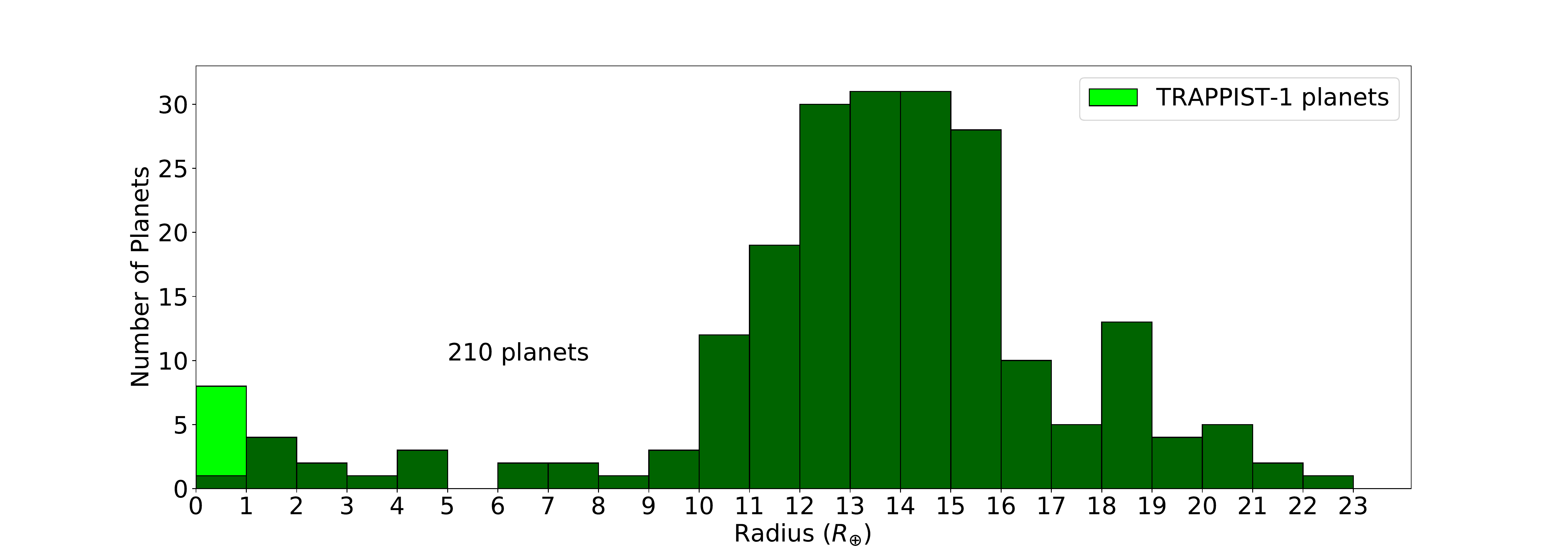}
\caption{ARIEL MRS with currently available planets radius distribution.}
\label{fig:known_radius}
\end{figure}

\begin{figure}[!htbp]
\centering
\includegraphics[scale=0.22]{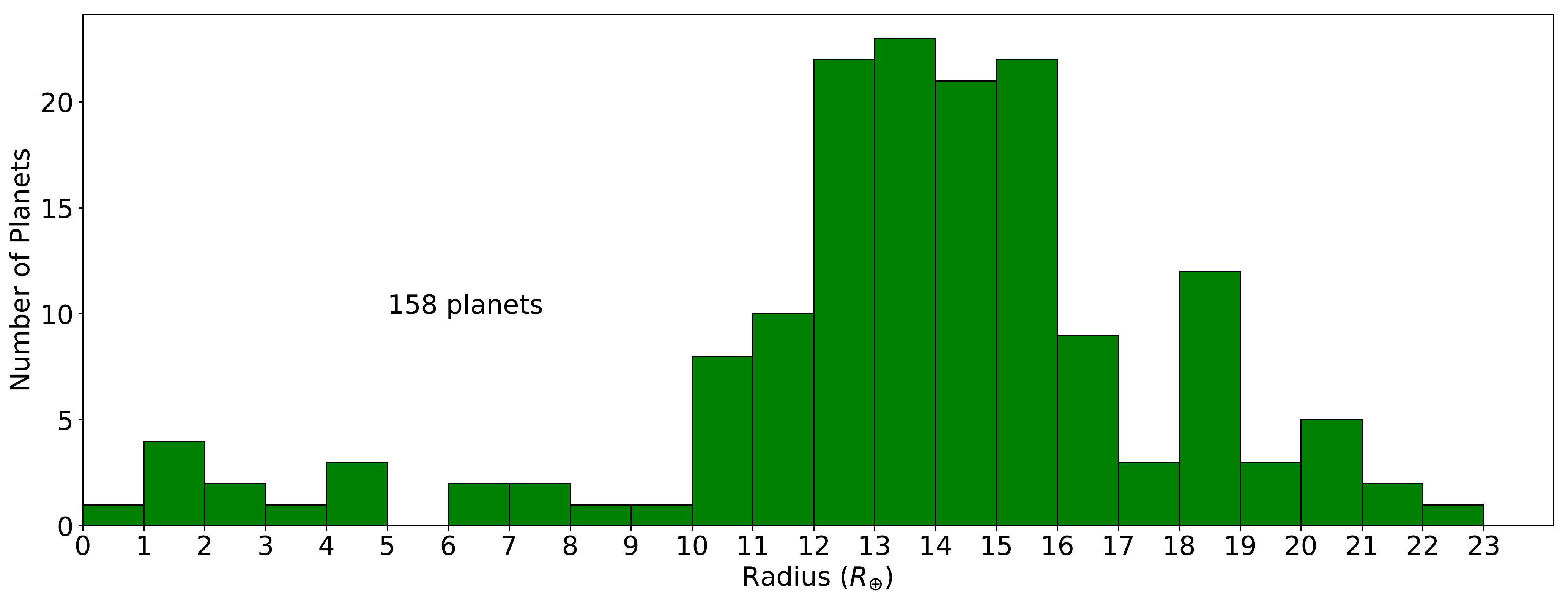}
\includegraphics[scale=0.22]{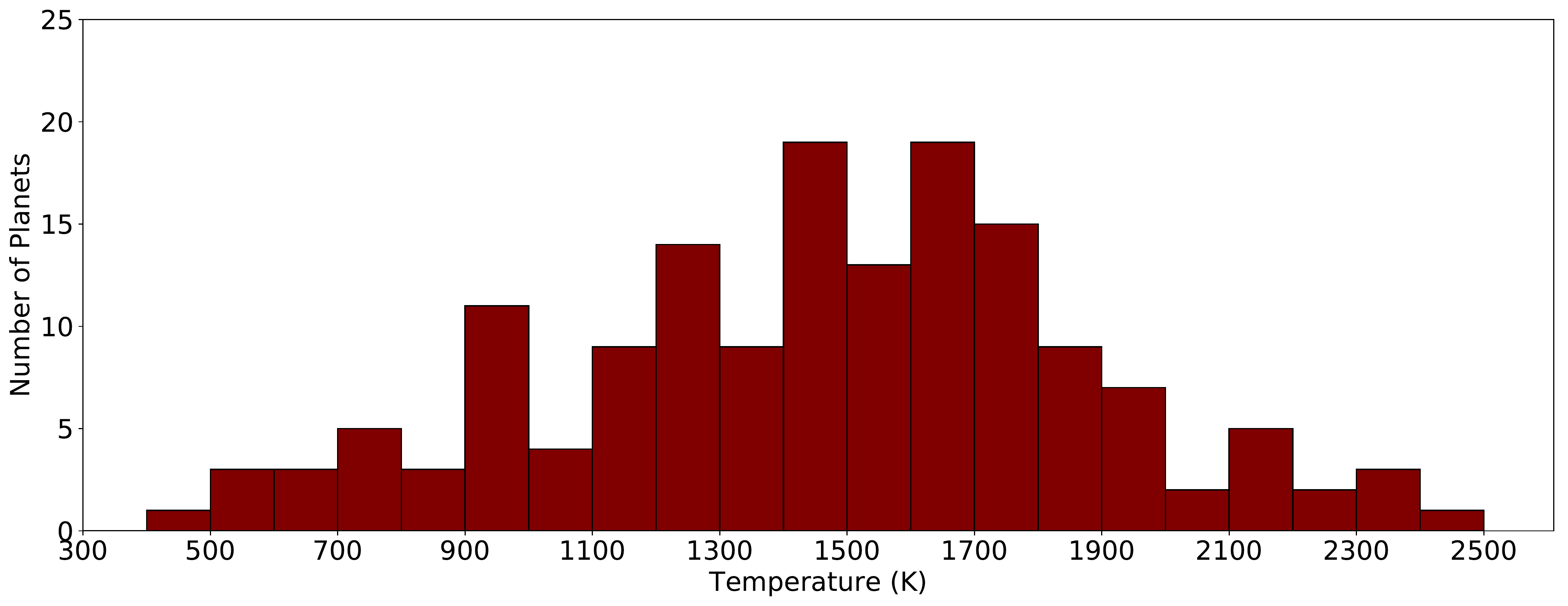}
\caption{Planets known today and observable by ARIEL in Deep mode, distributed in size-bins (\textbf{top}) and temperature bins (\textbf{bottom}) -- 158 planets.}
\label{fig:known_deep_radius}
\end{figure}

\begin{figure}[!htbp]
\centering
\includegraphics[scale=0.22]{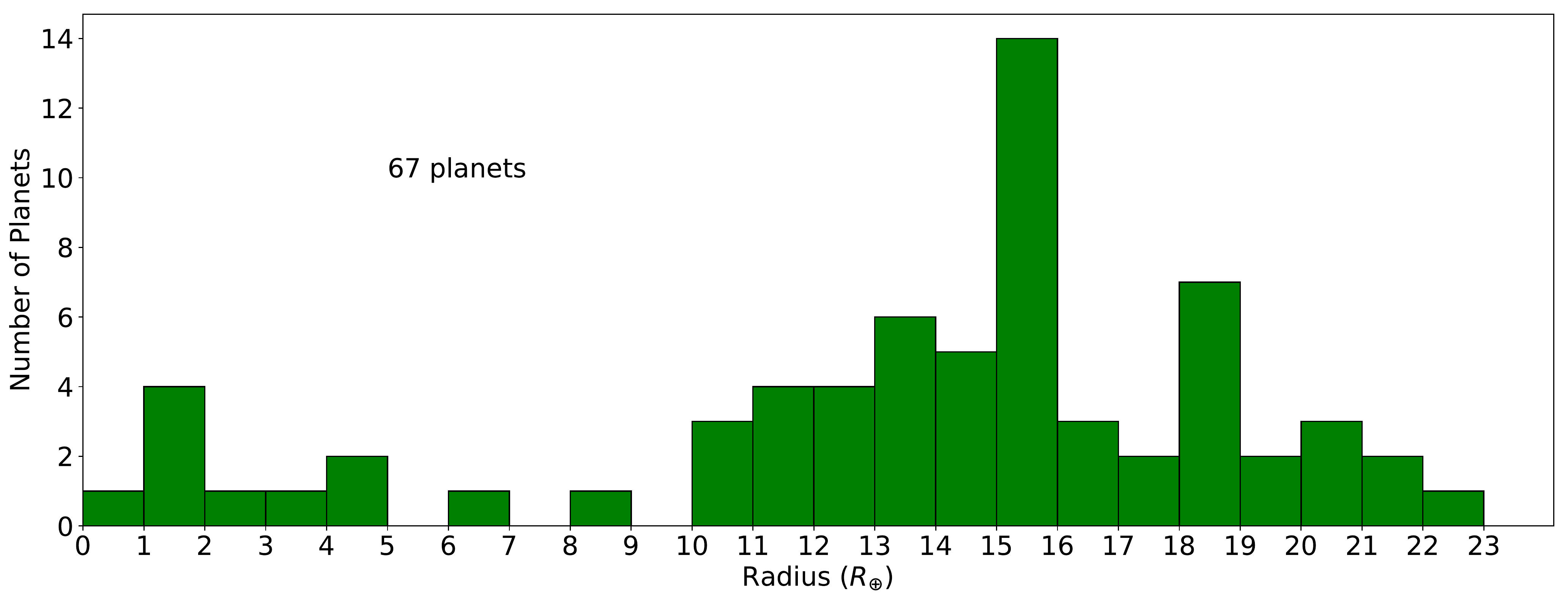}
\includegraphics[scale=0.22]{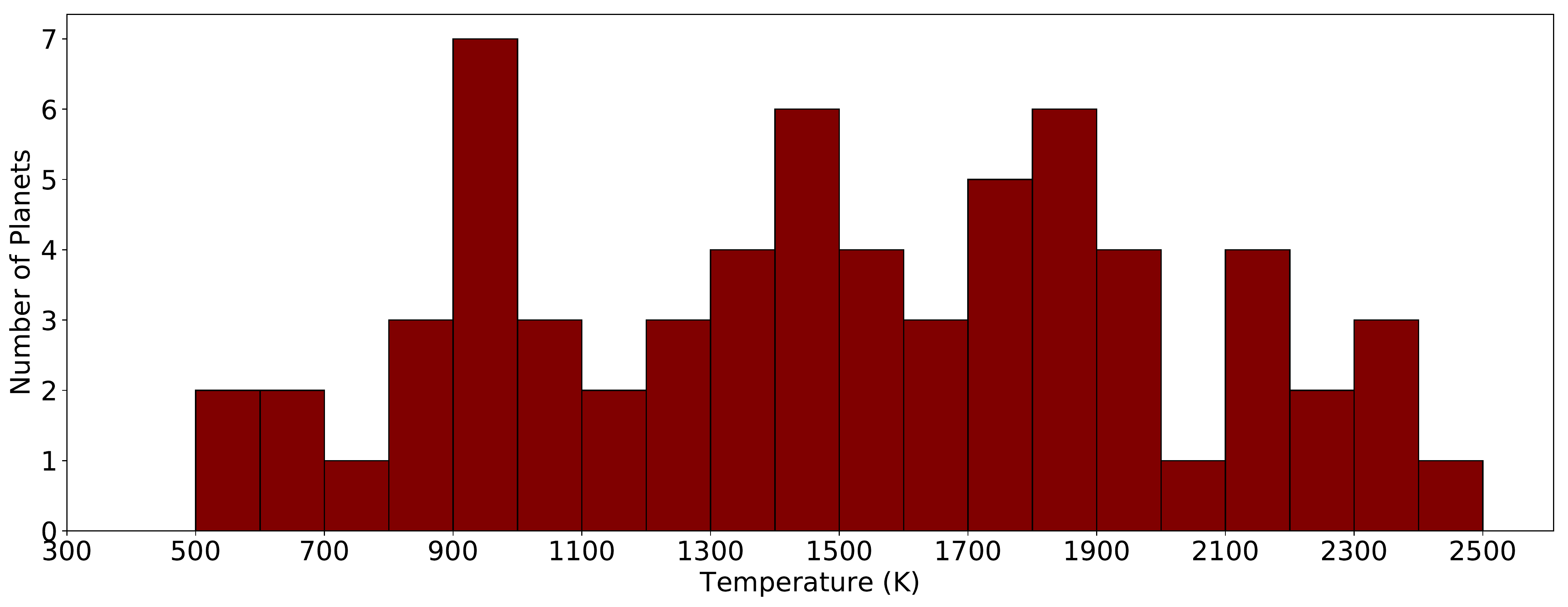}
\caption{Planets known today and observable by ARIEL in Benchmark mode, distributed size-bins (\textbf{top}) and temperature bins (\textbf{bottom}) -- 67 planets.}
\label{fig:known_benchmark_radius}
\end{figure}

In February 2017 $\sim$210 transiting planets fulfill the ARIEL previous criteria. It means that, even if ARIEL were launched tomorrow, it would observe at least 210 relevant targets.  Using the planets known today, we could organise the MRS into the following three tiers:
\begin{itemize}
\item Survey: 210 planets using 30\% of the mission lifetime (Fig \ref{fig:known_radius});
\item Deep: 158 planets using 60\% of the mission lifetime (Fig \ref{fig:known_deep_radius});
\item Benchmark: 67 planets using 10\% of the mission lifetime (Fig \ref{fig:known_benchmark_radius}).
\end{itemize}

\begin{figure}[!htbp]
\centering
\includegraphics[scale=0.22]{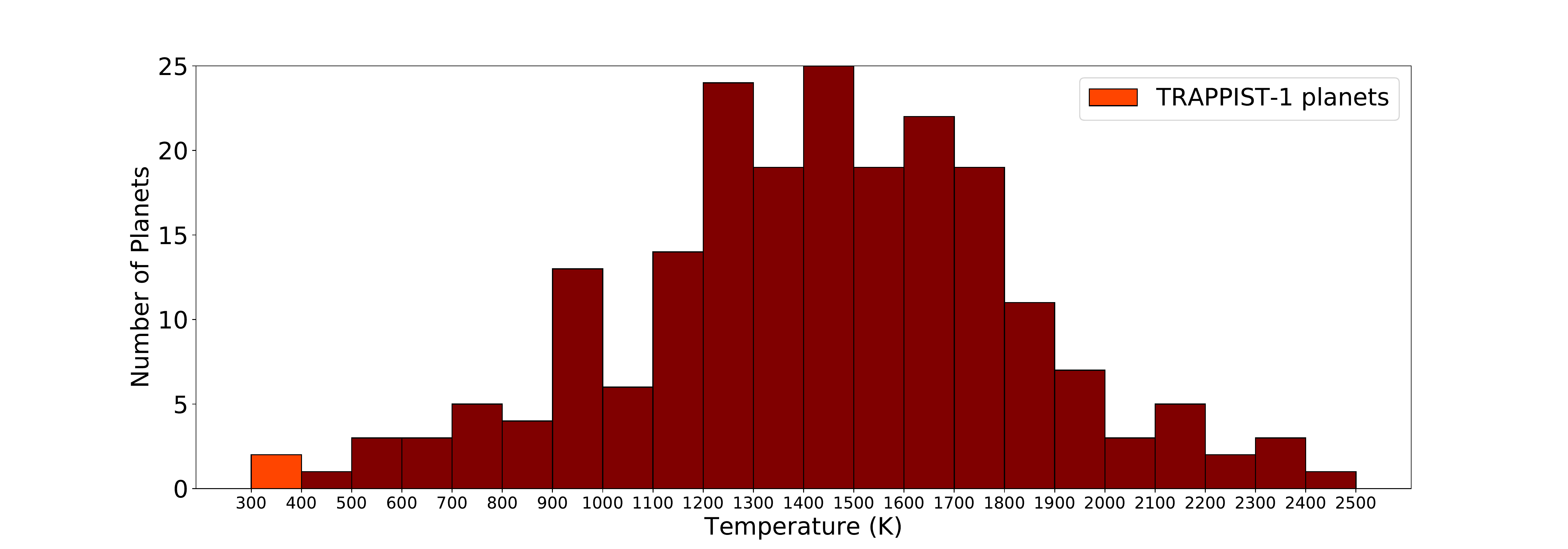}
\caption{ARIEL MRS with currently available planets temperature distribution.}
\label{fig:known_temperature}
\end{figure}

\begin{figure}[!htbp]
\centering
\includegraphics[scale=0.22]{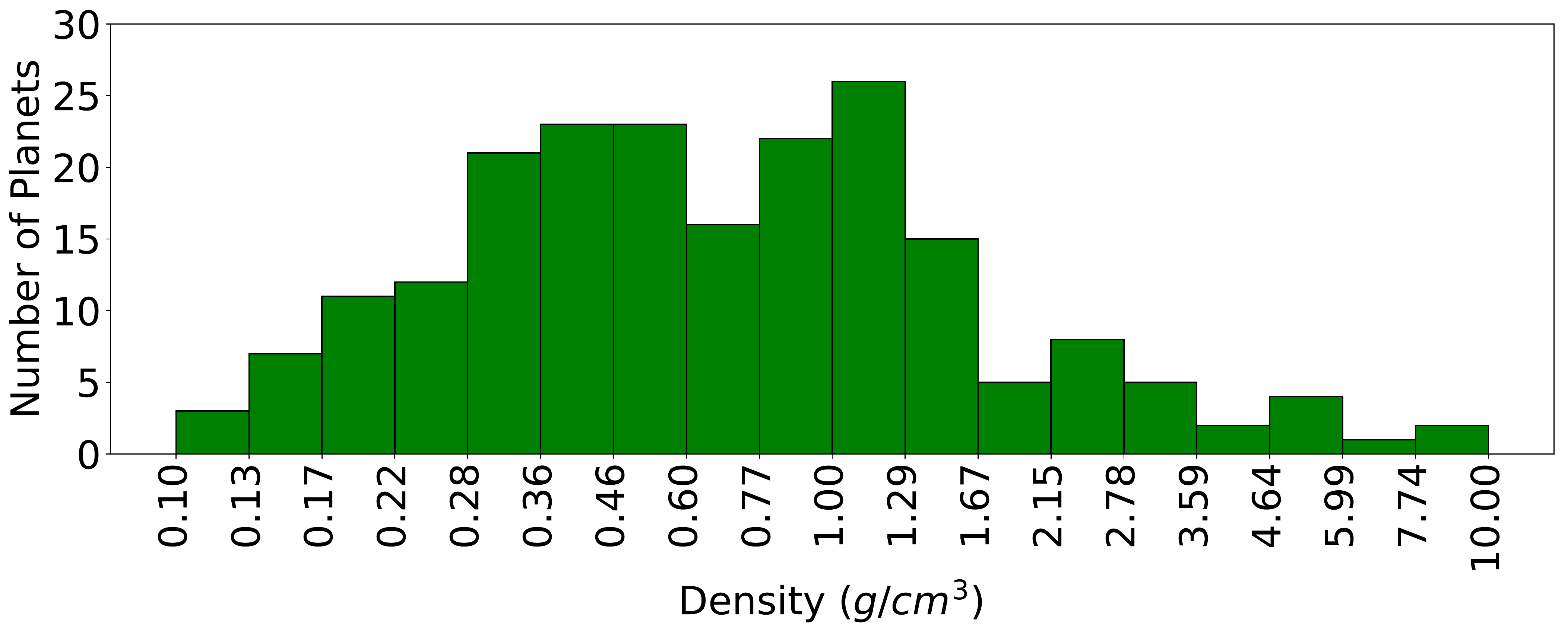}
\caption{ARIEL MRS with currently available planets density distribution.}
\label{fig:known_density}
\end{figure}

\begin{figure}[!htbp]
\centering
\includegraphics[scale=0.22]{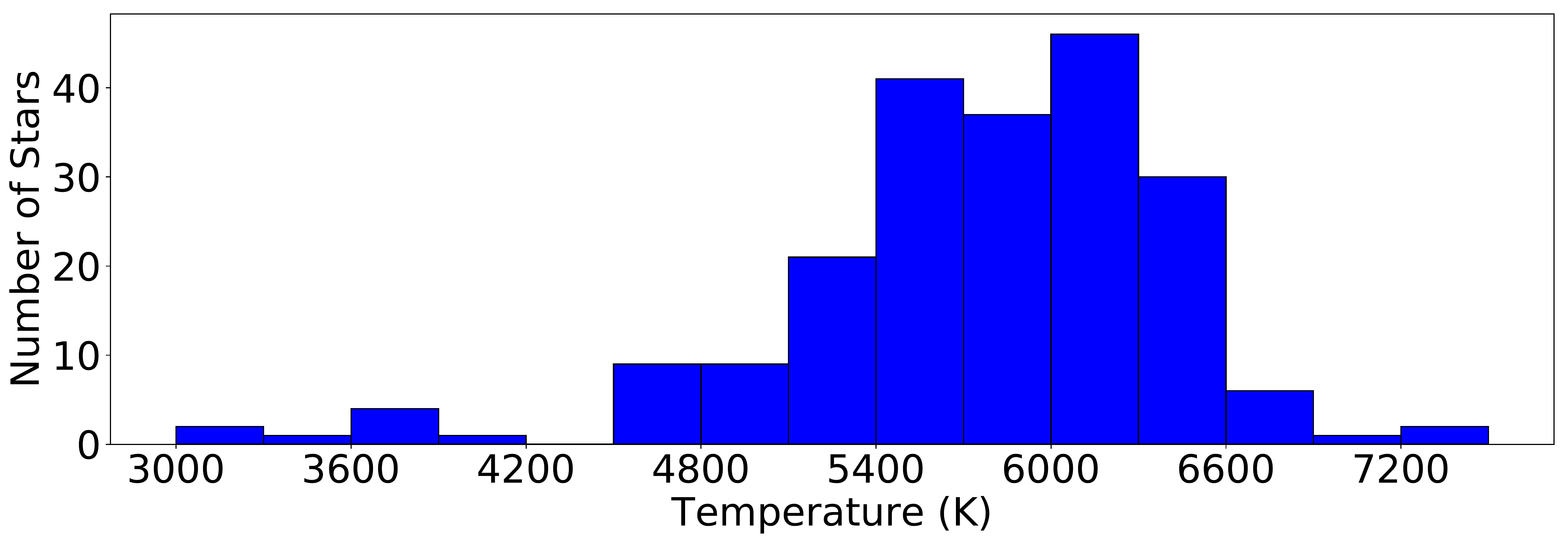}
\caption{Temperature distribution of the stellar hosts for the planets shown in fig. \ref{fig:known_radius}}
\label{fig:known_star_temperature}
\end{figure}

\begin{figure}[!htbp]
\centering
\includegraphics[scale=0.22]{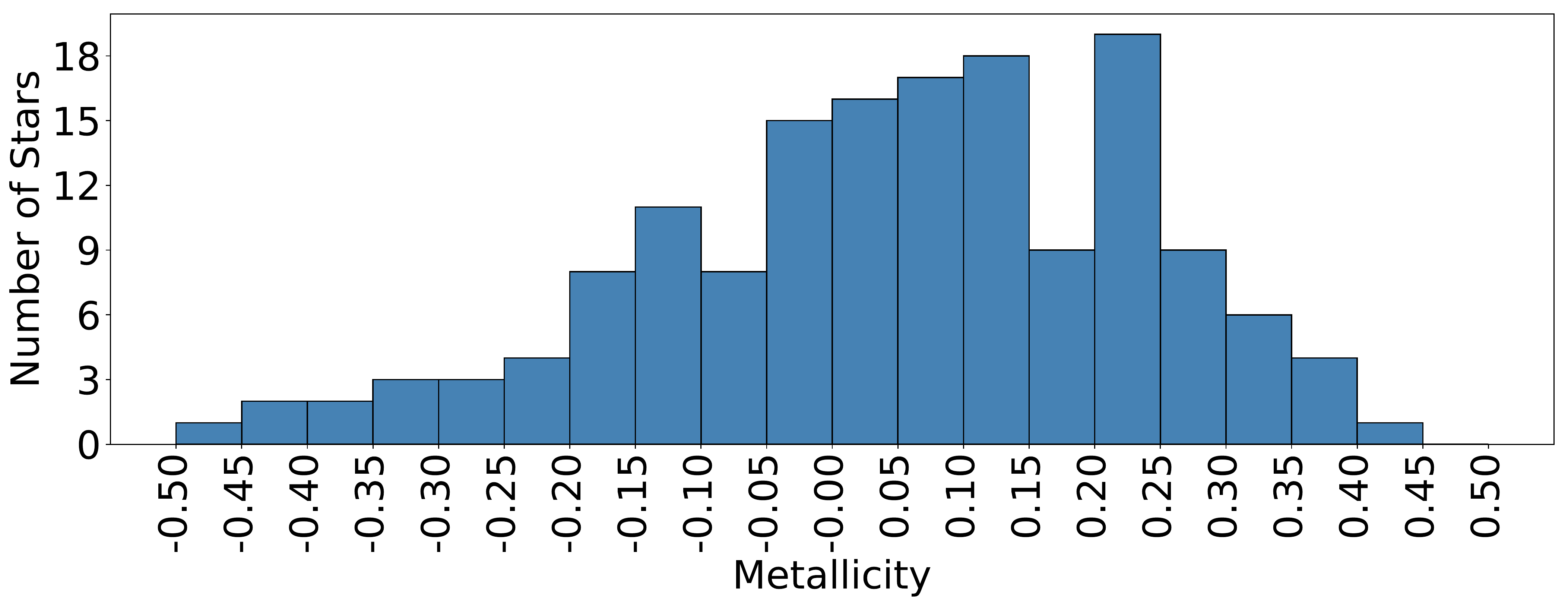}
\caption{Metallicity distribution of the stellar hosts for the planets shown in fig. \ref{fig:known_radius}}
\label{fig:known_star_metallicity}
\end{figure}

In Fig \ref{fig:known_radius}, \ref{fig:known_temperature} and \ref{fig:known_density} we show  the key physical parameters of the known planets defining the current observable MRS. In Fig \ref{fig:known_star_temperature} and \ref{fig:known_star_metallicity} we show the properties of the stellar hosts. As mentioned previously, the number of known planets is expected to increase dramatically in the future.

\begin{figure}[!htbp]
\centering
\includegraphics[scale=0.4]{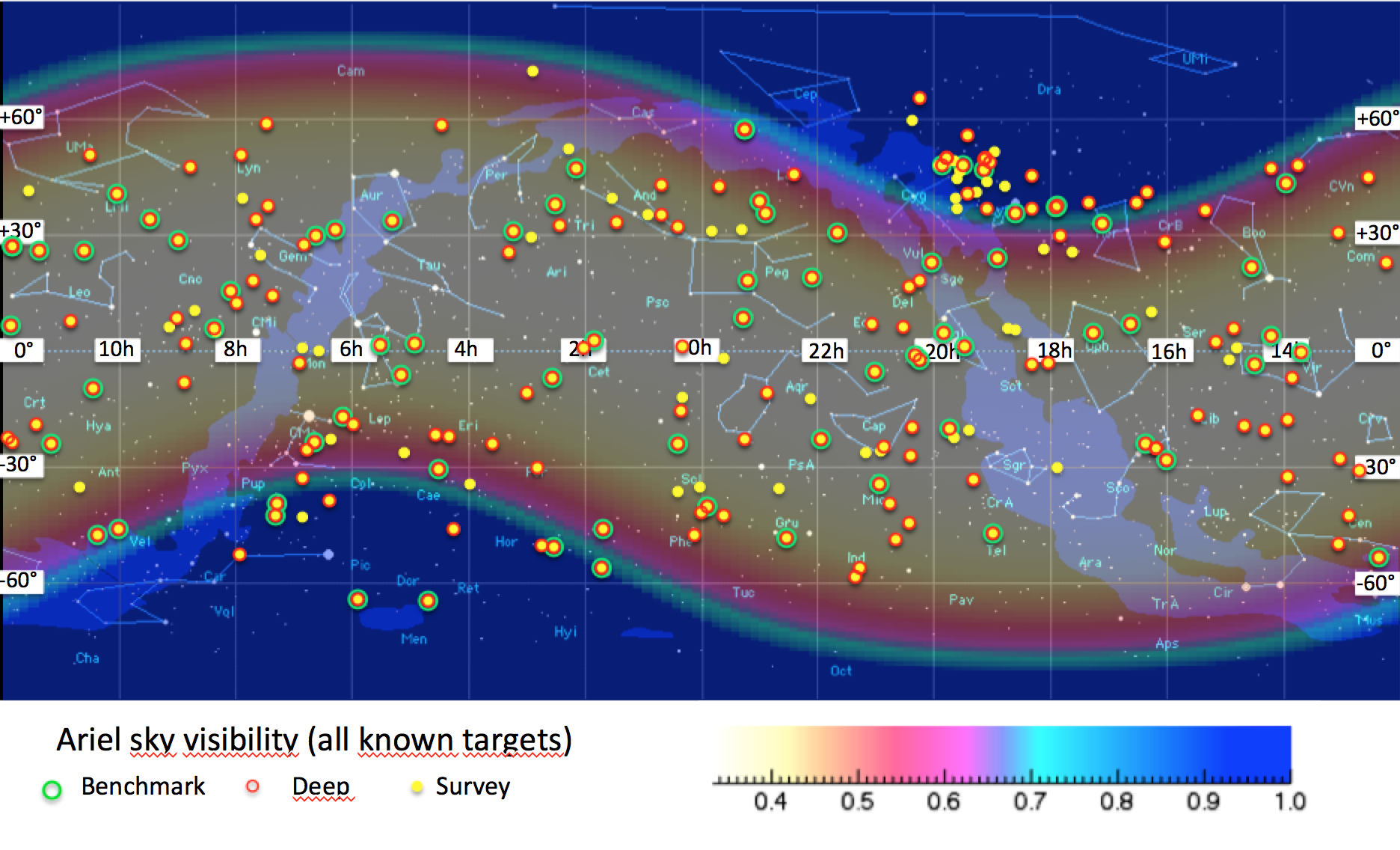}
\caption{A plot illustrating the fraction of the year for which a given location in the sky (in equatorial coordinates) is visible to ARIEL, as seen from a representative operational orbit of ARIEL at L2. Yellow dots: planets observed in Tier 1. Red dots: planets observed in Tier 2. Green dots: planets observed in Tier 3. (Marc Ollivier, private communication)}
\label{fig:ariel_sky_visibility}
\end{figure}

Pictorial representation (M. Ollivier, private comm.) of the known planets sky coordinates and their sky visibility all over the year is given in Fig \ref{fig:ariel_sky_visibility}. It shows that objects far away from the ecliptic plane will be visible longer than the planet close to this plane.

\clearpage

\section{MRS optimisation for stellar hosts}
\label{sec:other_method}
In this section we show another possible selection of the Tier 1 sample that maximises also the diversity of stellar hosts, additionally to other planet parameters. 
In particular, the stellar metallicity is expected to play an important role in the planet formation process and type of chemistry of the planet \citep{2015A&A...577A..33V}. ARIEL will also collect important data to understand better the relationship between stellar metallicity and planetary characteristics.

\subsection{Method}
We will limit our analysis to those systems which can be studied in up to six visits for each planet (either a transit or an occultation).

We chose four physical quantities that define a 4D space to distribute the ARIEL targets. The quantities are: stellar effective temperature (T$_\mathrm{eff}$ ), metallicity ([Fe/H]), planetary radius (R$_\mathrm{pl}$) and planetary theoretical equilibrium temperature (T$_\mathrm{pl}$). For the metallicity we use the values observed in the solar neighbourhood and reported by \citet{2011A&A...530A.138C}. We adopt three bins for stellar T$_\mathrm{eff}$, [Fe/H] and planetary R$_\mathrm{pl}$, while for the T$_\mathrm{pl}$ we use five bins, as detailed in Table \ref{tab_1}.
The three T$_\mathrm{eff}$ bins correspond approximately to the ranges of spectral types M-Late / K stars, Early K-G stars and F-G stars, respectively, as indicated in the labels in Fig \ref{fig:1002_planets} to \ref{fig:average_visits}. Analogously, we separated the sample in low metallicity, solar metallicity and high metallicity, according to individual temperature values. The binning into 3 intervals of T$_\mathrm{eff}$, [Fe/H] and R$_\mathrm{pl}$ is a reasonable trade-off between a detailed representation of the sample and a simple visualization of the richness and diversity of the physical configurations of the sample. We inferred from \citet{2011A&A...530A.138C} that the metallicities of stars in the solar neighbourhood are consistent with a normal distribution with mean -0.1 and standard deviation sd=0.2. Using such model distribution we simulated the values of [Fe/H] for each star in the ARIEL sample.

The 4D space of T$_\mathrm{eff}$ , [Fe/H], R$_\mathrm{pl}$ and T$_\mathrm{pl}$ is composed by a total of $3\times3\times3\times5=135$ cells. We assume that 10 systems are sufficiently reliable to determine the properties of the atmospheres of planets
in each cell.

\begin{table}
\caption{\label{tab_1} Bins of T$_\mathrm{eff}$, [Fe/H], R$_\mathrm{pl}$, T$_\mathrm{pl}$ defining the 4D parameter space.}
\centering
\resizebox{0.99\textwidth}{!}{ 
  \begin{tabular}{l|ccc}\hline\hline \\
  Stellar Temp.: T$_\mathrm{eff}$ & $3000 < T (\mathrm{K})< 4100$ & $4100 < T (\mathrm{K})< 5800$ & $T> 5800 \mathrm{K}$ \\
  Labels & M-Late K & Early K-G & F-G \\ \hline
  Metallicity: [Fe/H] & [Fe/H] $<$ -0.15&$-0.15 <$[Fe/H]$< 0.15$ & [Fe/H]$>0.15$ \\
  Labels & Low [Fe/H]& Solar & High [Fe/H] \\\hline
  Planet Radius: R$_\mathrm{pl}$ &R$_\mathrm{pl}<3 \mathrm{R_\oplus}$ & $3<\mathrm{R_\oplus}<8$ & R$_\mathrm{pl}>8 \mathrm{R_\oplus}$  \\
  Labels & Earths/ Super Earths& Neptunes & Jupiters \\\hline
  Planet Temp.: T$_\mathrm{pl}$ & \multicolumn{3}{c}{contiguous bins: [250, 500, 800, 1200, 1600, 2600] K} \\
  \hline
  \end{tabular}
}
\end{table}
\subsection{Results}
\begin{figure*}
	\centering
	\resizebox{0.99\textwidth}{!}{
		\includegraphics{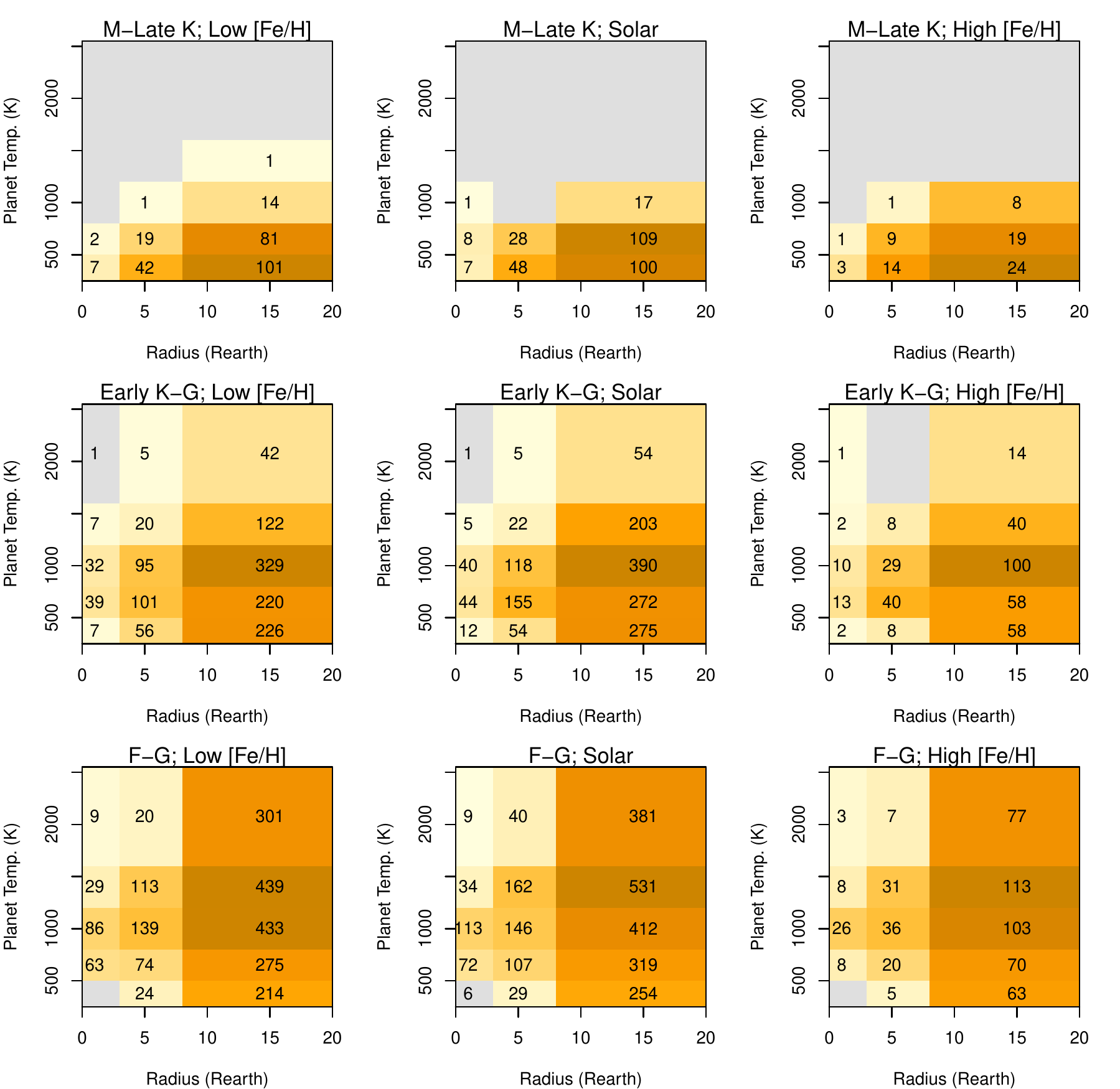}
	}
	\caption{\label{fig:9545_planets} Distribution of the 9545 planets in the 4D space of T$_\mathrm{eff}$, [Fe/H], R$_\mathrm{pl}$, T$_\mathrm{pl}$. Above each panels we indicate the spectral type and  metallicity. The numbers in each cell are the numbers of planets with the corresponding properties. The colour scale indicates more populated cells (darker orange/brown). Grey cells without any number indicate no objects.}
\end{figure*}

%Our ``population" thus contains a total of 9545 systems observable with up to 6 visits (see Fig. \ref{fig:Survey_all planets}), and including 210 known exoplanets. From this sample of 9545 systems it is possible to select the targets suitable for the ARIEL Tier 1 sample. 
The population of 9545 planets is distributed in the 4-D bins as in Fig \ref{fig:9545_planets}.

\begin{figure*}
	\centering
	\resizebox{0.99\textwidth}{!}{
		\includegraphics{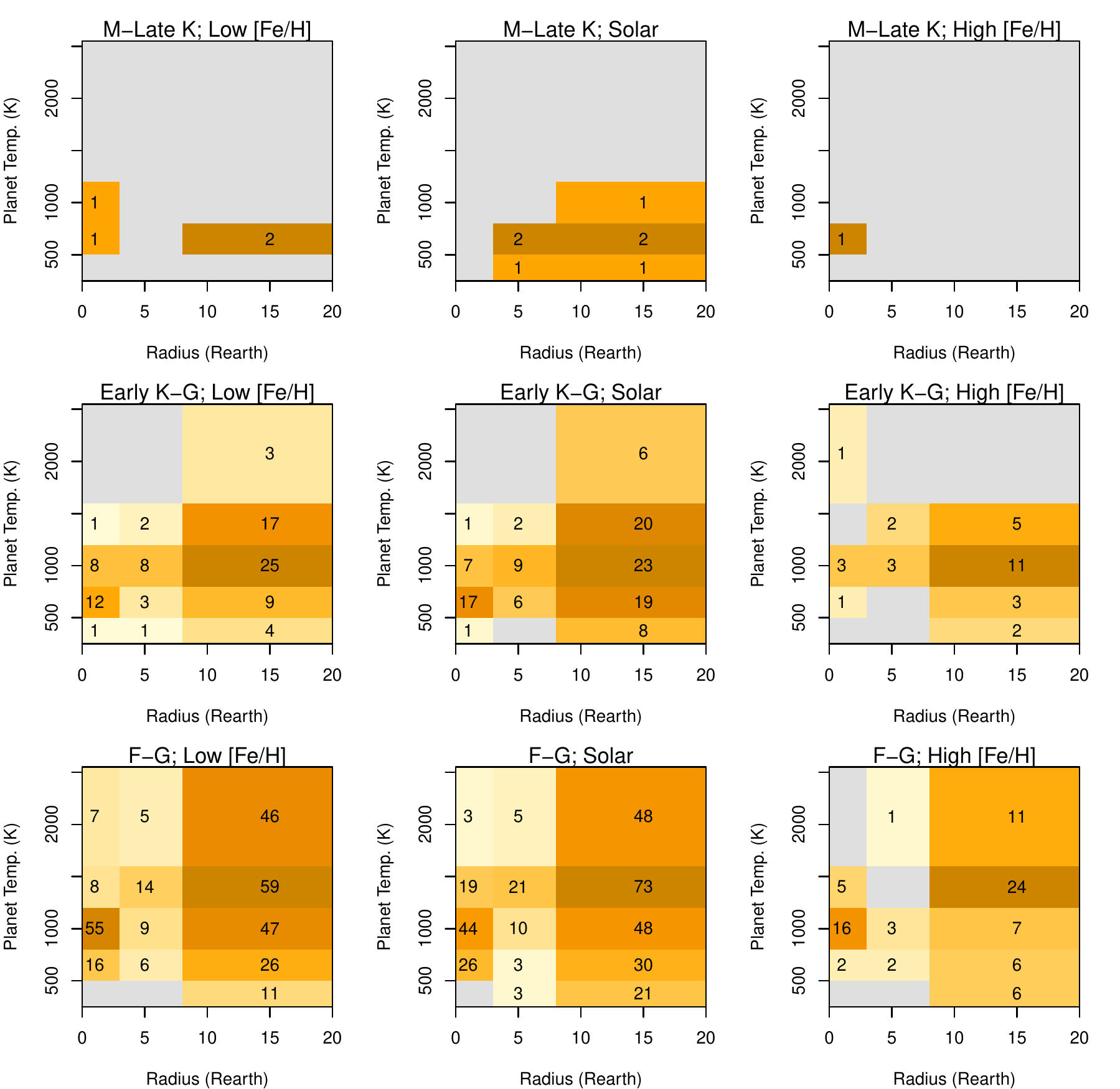}
	}
	\caption{\label{fig:1002_planets} Same as Fig. \ref{fig:9545_planets} 1002 planets of the Mission Reference Sample. }
\end{figure*}

From this distribution we selected 1002 exoplanets, requiring altogether 1538 satellite visits. These 1002 planets are distributed in the 4D space as shown in Fig. \ref{fig:1002_planets}. The $3\times3$ panel grid distributes the sample along the 3 spectral types and the metallicity ranges reported in Table \ref{tab_1}. Each panel is a matrix with planetary radii along x-axis and (calculated) equilibrium temperatures along y-axis, as specified in Table \ref{tab_1} and discussed above. The numbers in each box identify the numbers of systems with the corresponding R$_\mathrm{pl}$, T$_\mathrm{pl}$, spectral type, and [Fe/H] values.

\begin{figure*}
\centering
\resizebox{0.99\textwidth}{!}{
     \includegraphics{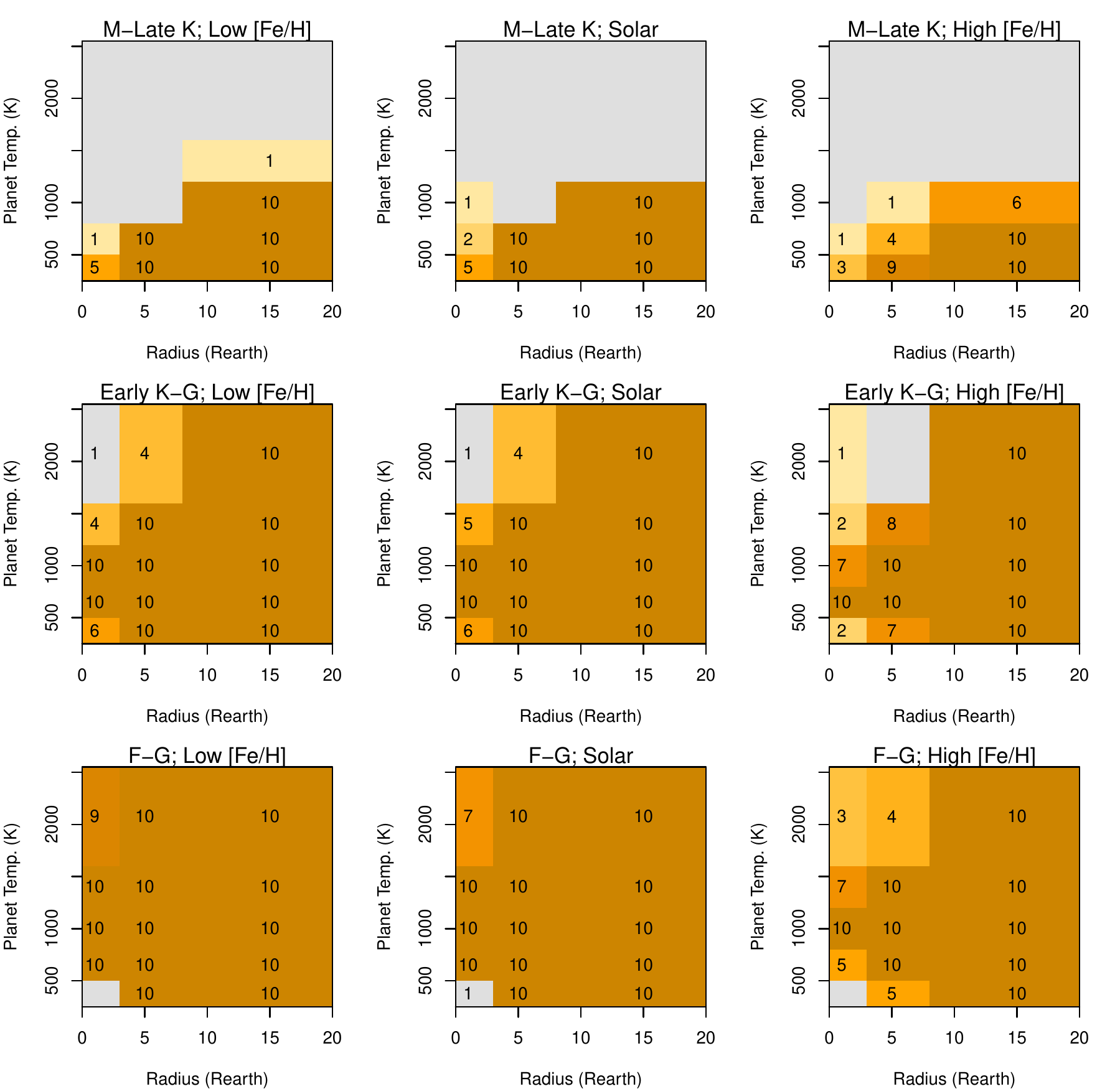}
     }
     \caption{\label{fig:908_planets} Same as Fig. \ref{fig:1002_planets} for the selected sample of 908 known and simulated planetary systems. They have been selected by filling each cell with up to 10 objects and for a budget of total satellite visits of about 1500.}
\end{figure*}

The 1002 systems in Fig. \ref{fig:1002_planets} tend to populate the cells corresponding to F-G-early and K stars orbited by Neptunes/Jupiters size planets (with a number of planets per cell $N>20$), as these systems are the easiest to be observed with high signal to noise and, on average, with one or two visits. At the same time, planets around M or late K stars are much less represented in this distribution, especially planets smaller than Neptunes. 
This issue is addressed by prioritising these targets over the rest of the population. We found that planets around M stars require on average more visits than the analogues around early K, G, and F stars. We managed to select 908 planets and, in particular, 594 of them require only 1 visit (65.4\%), 151 planets require 2 visits (16.6\%), 83 planets require 3 visits (9.1\%), 41 planets require 4 visits (4.5\%), and 39 planets require 5 visits (4.4\%).
The corrected sample is shown in Fig \ref{fig:908_planets}, where now $\sim 19$\% of the population are Earths/Super Earth or Neptunes around M or K stars observable with less than 6 visits.

Assuming a total number of visits as in the 1002 planets configuration (approximately 1500 visits), we fixed the maximum number of systems (10 planets in our choice) in each 4D space cell. This choice implies that any additional targets in an ``already full" cell will be discarded. In this way we can include planets in the empty or poorly populated parts of the parameter space. The goal is to verify that we can cover with enough statistics most of the 4D parameter space. The distribution of systems selected with such criteria is shown in Fig. \ref{fig:908_planets}. Compared to Fig. \ref{fig:1002_planets}, we see that we can efficiently cover most of the 4D space in planetary sizes, planetary temperatures, host temperatures and metallicities, apart from those combination of parameters corresponding to not physical or rare systems (e.g., very hot planets around very cool stars). Our selection is composed by 908 unique planets requiring a total of 1504 visits. Among already known systems, 92 of the initial 211 systems are in this new list. This selection is not unique, and depends on our choices, but our exercise shows that we have great freedom on the final choice on how to spend ARIEL observing time, as it can be easily tuned on specific needs.
\begin{figure*}
\centering
\resizebox{0.99\textwidth}{!}{
     \includegraphics{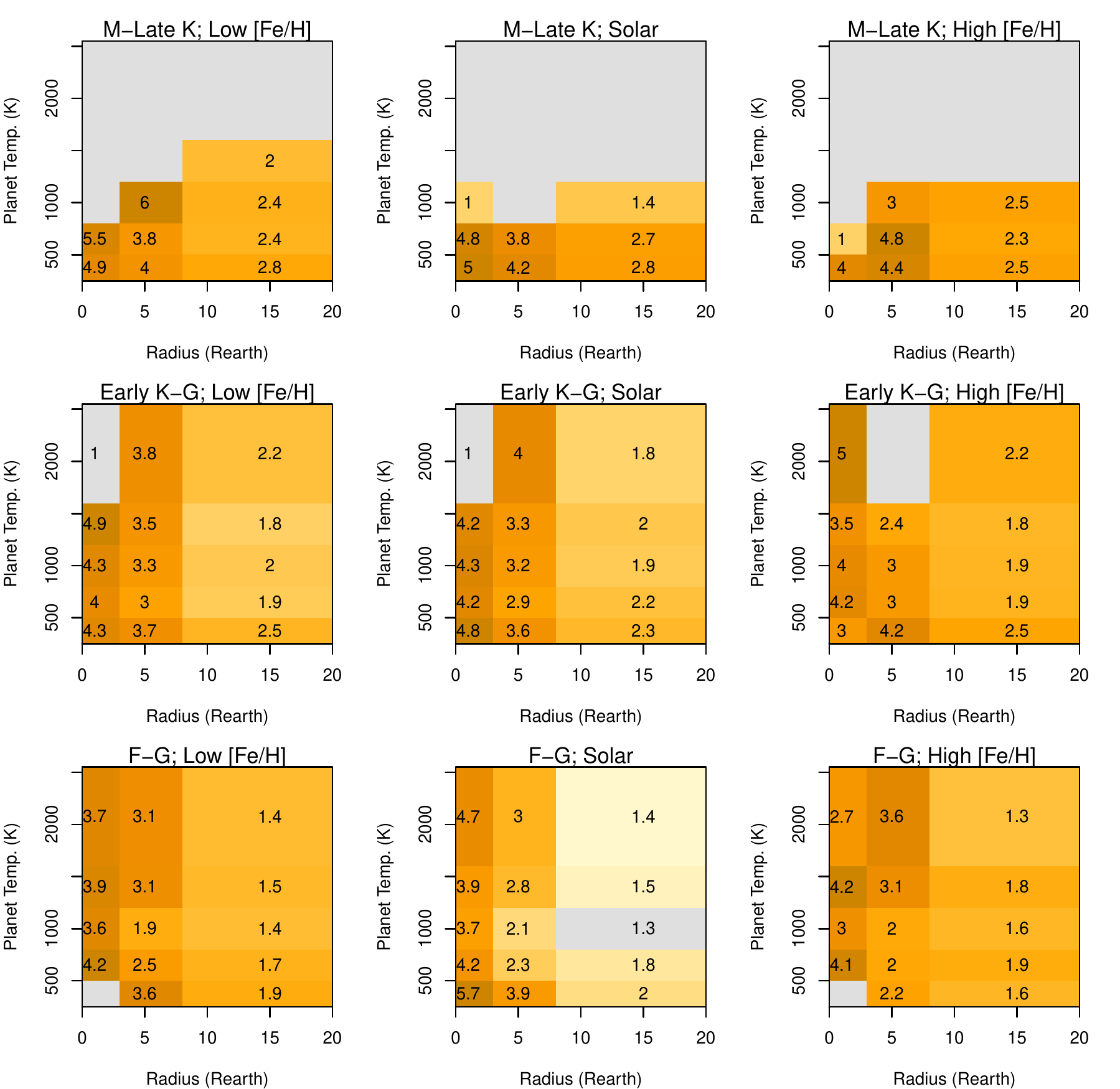}
     }
     \caption{\label{fig:average_visits} Average number of visits required for the sample selected in Fig. \ref{fig:908_planets}. The binning is as in Figs. \ref{fig:1002_planets} to \ref{fig:908_planets}.}
\end{figure*}
Fig. \ref{fig:average_visits} shows the average number of visits required to cover each cell of the 4D space. The number of visits needed for Jupiters and Neptunes is, typically, one or two, while Earths/Super Earths require from 3 to 5 visits each. 
 To summarise, out of the 908 planets in our selection there are 594 planets requiring only 1 visit (65.4\%), 151 planets requiring 2 visits (16.6\%), 83 planets requiring 3 visits (9.1\%), 41 planets requiring 4 visits (4.5\%), and 39 planets requiring 5 visits (4.4\%).

As a final comment, we have verified that, by increasing the maximum number of systems per 4D cell while keeping fixed the total number of visits to $\sim1500$, we obtain that the number of observed planets increases (for example assuming N=15 as maximum systems per cell, we can observe up to 1000 systems), but at the same time the 4D cells of systems with cold/warm Earths/Super Earths would tend to be left empty and thus unexplored. This exercise shows the degree of flexibility offered by ARIEL in the choice of the target sample.

\section{Conclusions}\label{sec:conclusion}
In this paper we demonstrated that the current ARIEL design enables the observation of 900-1000 planets during its four-year lifetime, depending on the physical parameters of the planet/star systems which one wants to optimise.
The optimal sample of targets fulfils all the science objectives of the mission.
While we currently know only $\sim$200 transiting exoplanets which could be part of the mission reference sample, new space missions and ground-based observatories are expected to discover thousands of new planets in the next decade. NASA-TESS alone is expected to deliver most ARIEL targets.

\section*{Ackowledgements} 
T. Z. is supported by the European Research CouncilERC projects \emph{ExoLights} (617119) and from INAF trough the "Progetti Premiali” funding scheme of the Italian Ministry of Education, University, and Research. I.P and G.M. are supported by Ariel ASI-INAF agreement No. 2015-038-R.0. G.T. is supported by a Royal Society URF. 
We thank Enzo Pascale and Ludovig Puig for their help in setting up the ESA's Radiometric model.

\clearpage

\begin{appendices}
\section{ESA Radiometric Model validation with ExoSim}\label{app:validation}
We compare the out-of-transit signal and noise from ESA Radiometric Model (ERM) with that from ExoSim. An early version of ARIEL with a grating design was used for the instrument model in each.  We model 55 Cancri and GJ 1214 with the same PHOENIX spectra in each simulator and include only photon noise and the noise floor, $N_{min}(\lambda)$, which is dominated by dark current noise. All the calculations are done per unit time and per spectral bin ($R=30$ in Ch1 and $R=100$ in Ch0). The noise variance was compared assuming an aperture mask on the final images, and the noiseless signal per unit time was compared assuming no aperture. In the ERM, we use the following expression for $N_{min}$ giving the noise variance:
\begin{equation}
N_{min}(\lambda) = \frac{2.44 f\lambda^2}{mR{\Delta_{pix}}^2}I_{dc}
\end{equation}
where $I_{dc}$ is the dark current per pixel, $m$ is the reciprocal linear dispersion of the spectrum in $\mu$m wavelength per $\mu$m distance, $R$ is the spectral resolving power and $\Delta_{pix}$ is the pixel pitch.
The noise variance from ExoSim is given as the average of 50 realizations with a standard deviation (shown as error bars in the following figures).
For 55 Cancri e case (Fig \ref{fig:55cnce1}), over all wavelength bins, the ERM signal is always within 2\% of ExoSim, and the averaged noise variance within 5\% of the ERM.  In 94\% of the bins, the ERM noise variance is within the standard deviation from ExoSim.

\begin{figure}[!htbp]
\centering
\includegraphics[scale=0.16]{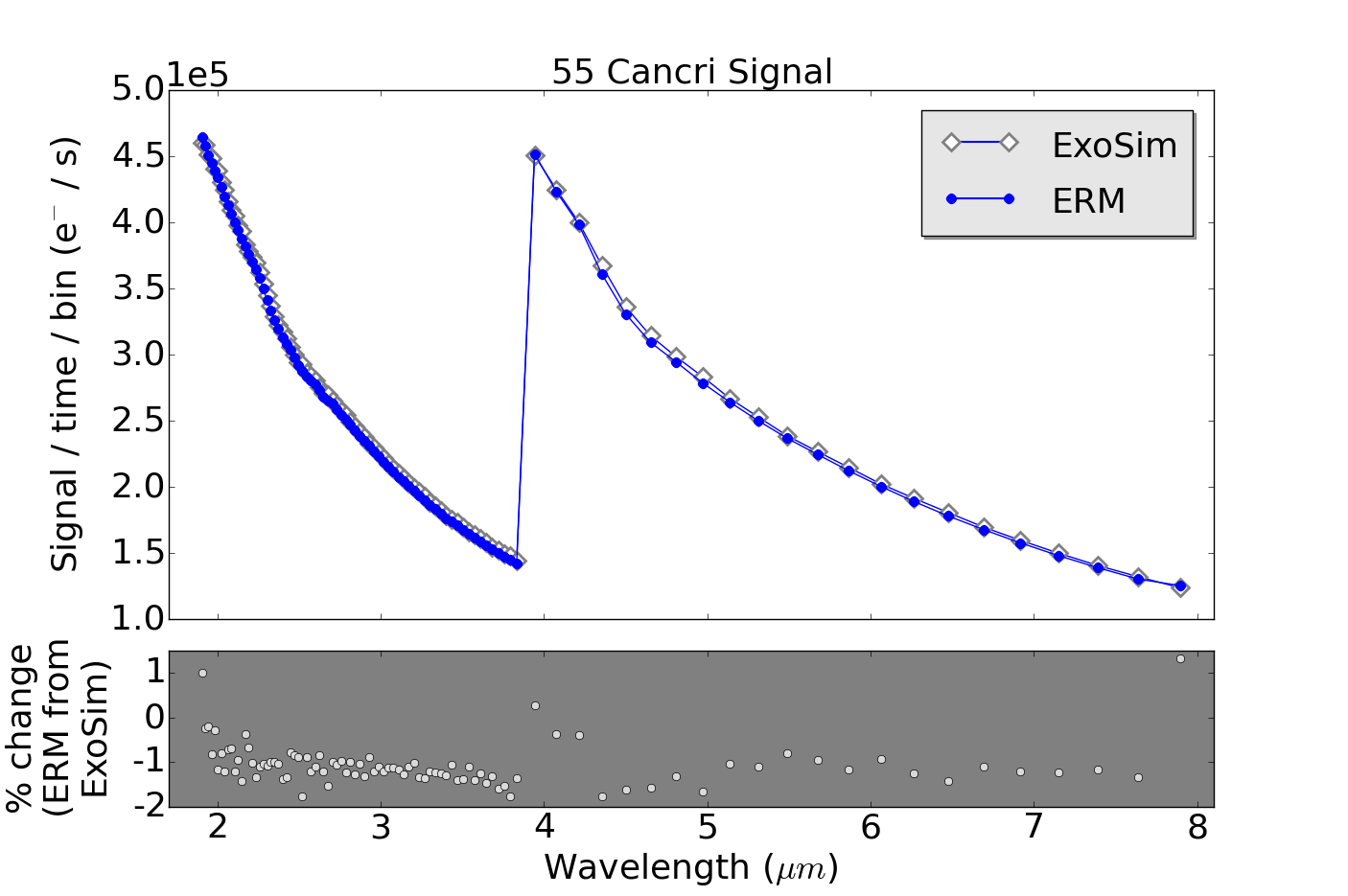}
\includegraphics[scale=0.16]{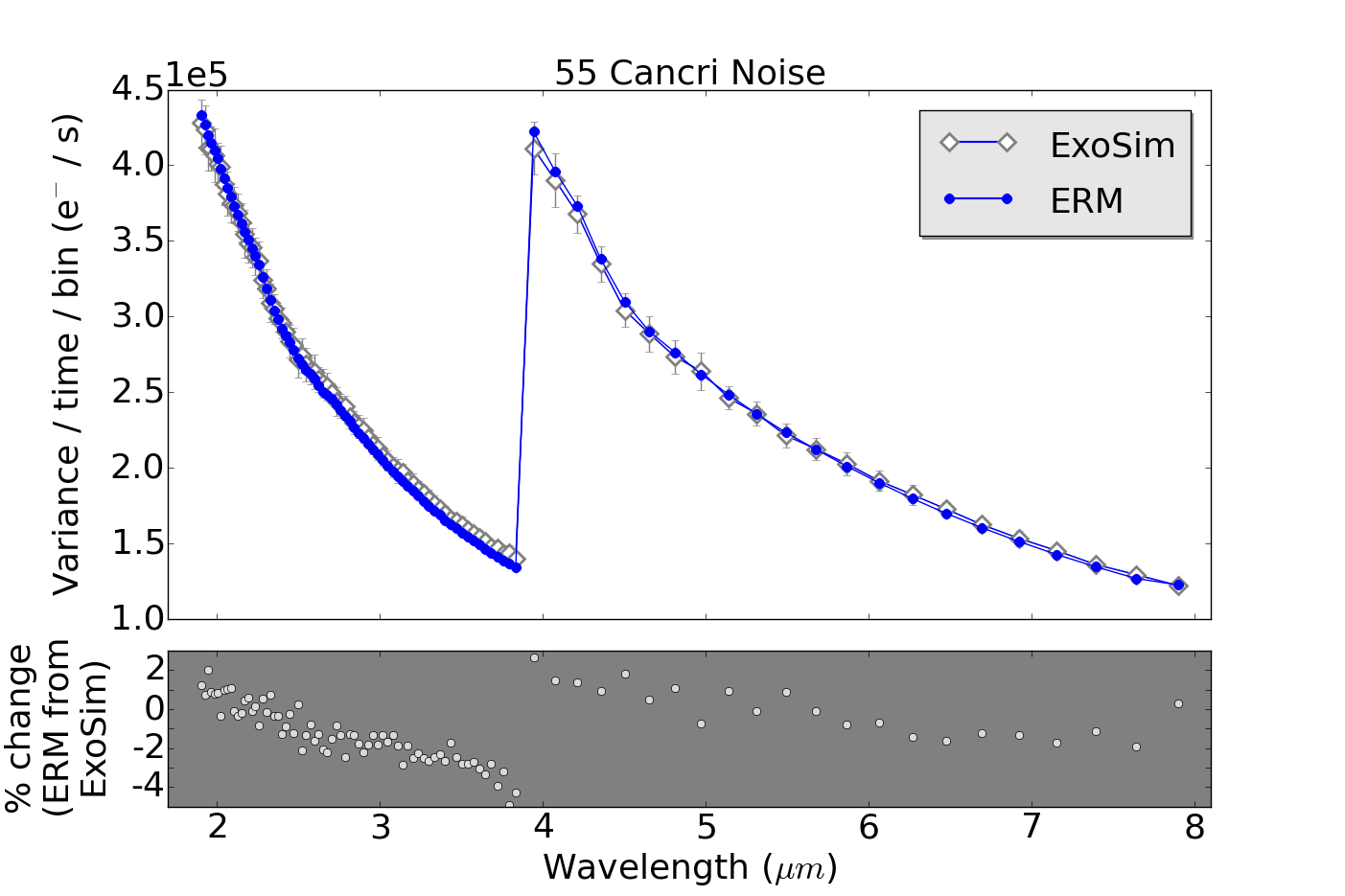}
\caption{Comparison between the out-of-transit signal (left) and noise (right) simulated by ExoSim (white points) and the ESA Radiometric Model (blue points) for the star 55 Cancri.  Subplots show the percent difference of the ERM from ExoSim.}
\label{fig:55cnce1}

\end{figure}

\begin{figure}[!htbp]
\centering
\includegraphics[scale=0.16]{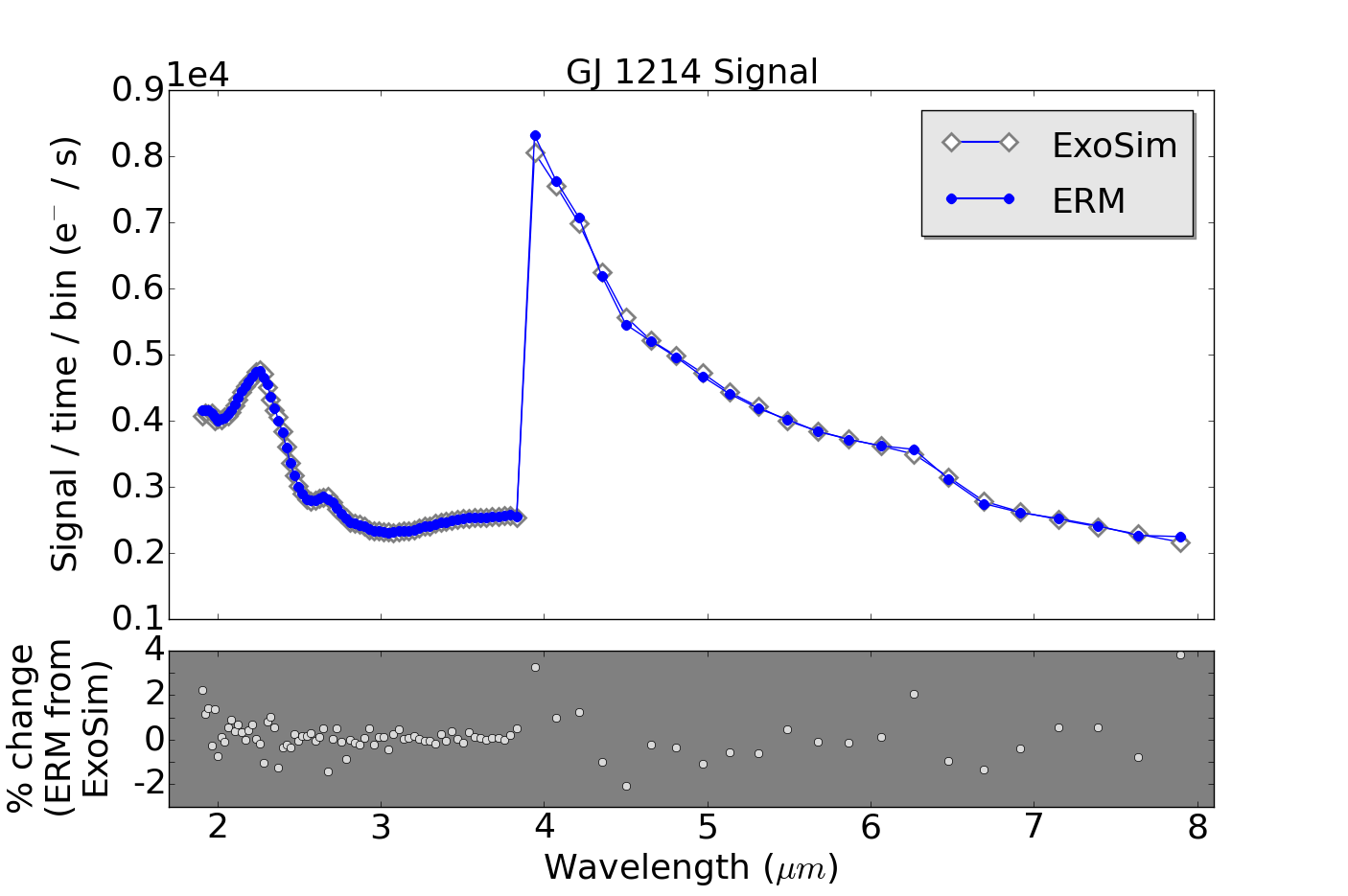}
\includegraphics[scale=0.16]{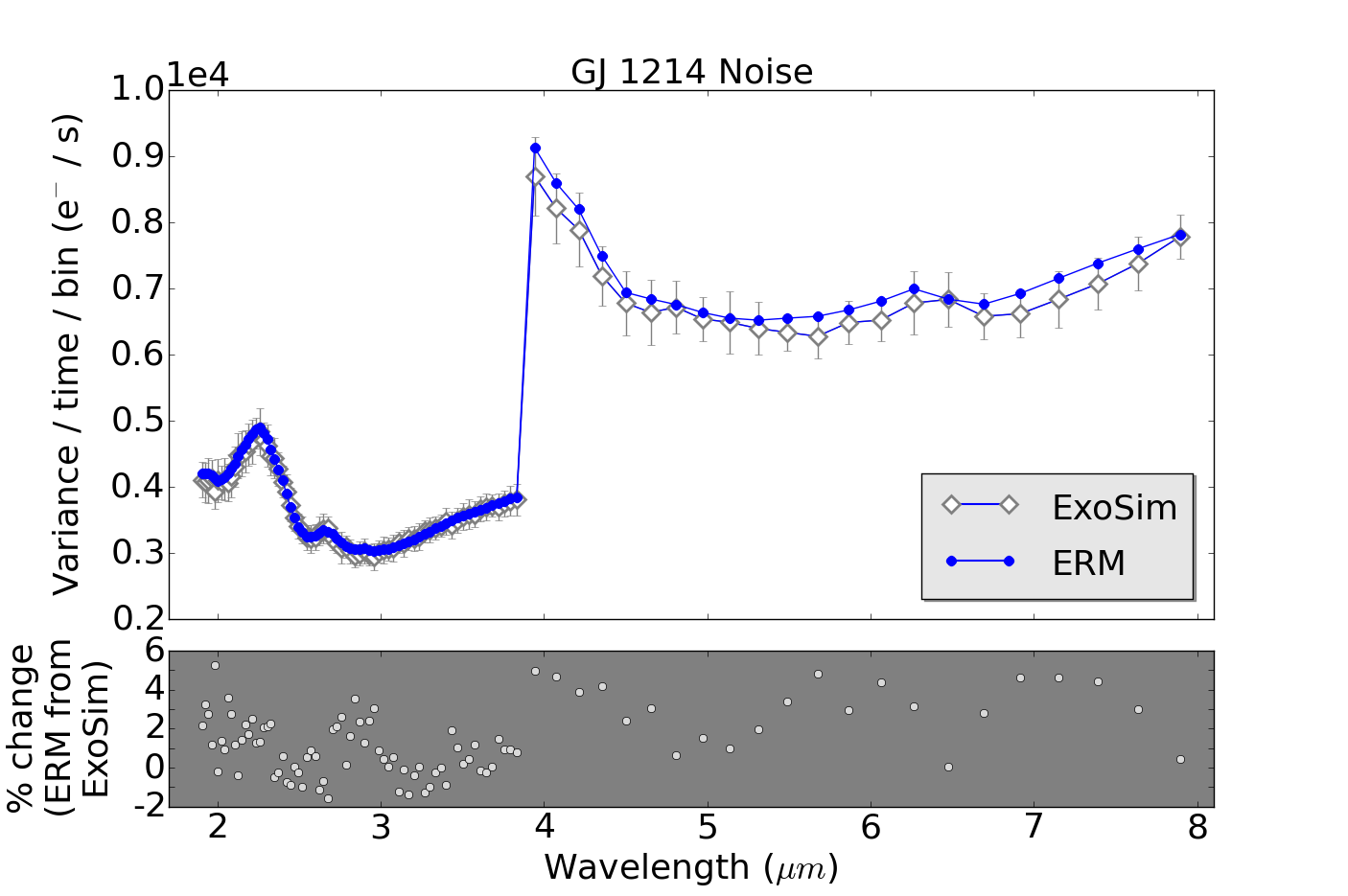}
\caption{Comparison between the out-of-transit signal (left) and noise (right) simulated by ExoSim (white points) and the ESA Radiometric Model (blue points) for the star GJ 1214. Subplots show the percent difference of the ERM from ExoSim.} \label{fig:gj1214b3}

\end{figure}

For GJ 1214 (Fig \ref{fig:gj1214b3}), the ERM signal is within 4\% of ExoSim over all bins and the averaged noise variance within 6\% of ExoSim over all bins. The ERM noise variance is always within the standard deviation from ExoSim over all bins.

There is therefore good agreement between the two simulators.

\clearpage

\section{Known planets observable by ARIEL} \label{app:known_planets}

\begin{table}[!ht]
\centering

\begin{small}
\resizebox{\columnwidth}{!}{%
\begin{tabular}{|c|c|c|c|c|c|c|c|c|c|c|}
\hline
\multirow{2}{*}{\#} & \multirow{2}{*}{\textbf{Planet}} &
\multicolumn{4}{c|}{\textbf{planetary properties}} &
\multicolumn{2}{c|}{\textbf{stellar properties}} &
\multicolumn{2}{c|}{\textbf{Observation}}\\
\cline{3-10}
  &  & { R $(R_{\oplus})$} &{\ M $(M_{\oplus})$} & {P (days)} & {T (K)} & { R $(R_{\odot}$)} &{ T (K)} & {\#} & {type} \\
    \hline
    1 & 55 Cnc e & 1.88 & 8.07 & 0.74 & 1891 & 0.95 & 5196 & 1 & transit \\ 
\hline
2 & \footnotesize{EPIC 204129699 b} & 15.47 & 563.97 & 1.26 & 1473 & 0.91 & 5280 & 1 & transit \\ 
\hline
3 & WASP-52 b & 13.94 & 146.24 & 1.75 & 1267 & 0.87 & 5000 & 1 & transit \\ 
\hline
4 & HD 189733 A b & 12.49 & 361.78 & 2.22 & 1180 & 0.80 & 4980 & 1 & transit \\ 
\hline
5 & WASP-77 A b & 13.28 & 559.52 & 1.36 & 1762 & 1.00 & 5500 & 1 & transit \\ 
\hline
6 & WASP-85 A b & 16.24 & 387.85 & 2.66 & 1341 & 1.04 & 5685 & 1 & transit \\ 
\hline
7 & WASP-33 b & 15.78 & 1459.19 & 1.22 & 2541 & 1.50 & 7200 & 1 & occultation \\ 
\hline
8 & WASP-19 b & 15.21 & 371.32 & 0.79 & 1998 & 0.97 & 5500 & 1 & occultation \\ 
\hline
9 & WASP-95 b & 13.28 & 359.23 & 2.18 & 1521 & 1.11 & 5630 & 1 & transit \\ 
\hline
10 & WASP-121 b & 19.83 & 376.08 & 1.27 & 2295 & 1.35 & 6459 & 1 & transit \\ 
\hline
11 & WASP-12 b & 19.05 & 446.34 & 1.09 & 2399 & 1.35 & 6118 & 1 & occultation \\ 
\hline
12 & WASP-35 b & 14.48 & 228.89 & 3.16 & 1414 & 1.07 & 5990 & 1 & transit \\ 
\hline
13 & HAT-P-30 b & 14.70 & 226.03 & 2.81 & 1594 & 1.24 & 6304 & 1 & transit \\ 
\hline
14 & WASP-108 b & 14.09 & 283.57 & 2.68 & 1558 & 1.17 & 6000 & 2 & transit \\ 
\hline
15 & HD 209458 b & 15.14 & 226.99 & 3.52 & 1401 & 1.15 & 6075 & 1 & transit \\ 
\hline
16 & WASP-122 b & 21.64 & 436.17 & 1.71 & 1900 & 1.40 & 5720 & 1 & transit \\ 
\hline
17 & WASP-2 A b & 12.26 & 290.57 & 2.15 & 1276 & 0.89 & 5255 & 2 & transit \\ 
\hline
18 & HAT-P-32 b & 22.35 & 299.15 & 2.15 & 1850 & 1.18 & 6207 & 1 & transit \\ 
\hline
19 & WASP-43 b & 11.37 & 646.62 & 0.81 & 1403 & 0.72 & 4520 & 1 & occultation \\ 
\hline
20 & WASP-123 b & 14.56 & 292.47 & 2.98 & 1477 & 1.21 & 5740 & 1 & transit \\ 
\hline
21 & WASP-101 b & 15.47 & 158.95 & 3.59 & 1518 & 1.34 & 6400 & 1 & transit \\ 
\hline
22 & WASP-74 b & 17.12 & 302.01 & 2.14 & 1872 & 1.48 & 5990 & 1 & transit \\ 
\hline
23 & WASP-76 b & 20.08 & 292.47 & 1.81 & 2125 & 1.46 & 6250 & 1 & transit \\ 
\hline
24 & WASP-1 b & 16.27 & 271.49 & 2.52 & 1777 & 1.24 & 6160 & 1 & occultation \\ 
\hline
25 & KELT-10 b & 15.35 & 215.86 & 4.17 & 1340 & 1.11 & 5948 & 1 & transit \\ 
\hline
26 & KELT-3 b & 14.90 & 464.78 & 2.70 & 1774 & 1.28 & 6304 & 1 & transit \\ 
\hline
27 & WASP-62 b & 15.25 & 181.21 & 4.41 & 1389 & 1.25 & 6230 & 1 & transit \\ 
\hline
28 & HD 149026 b & 7.88 & 113.17 & 2.88 & 1699 & 1.30 & 6147 & 1 & transit \\ 
\hline
29 & WASP-97 b & 12.40 & 419.64 & 2.07 & 1500 & 1.12 & 5640 & 2 & occultation \\ 
\hline
30 & WASP-94 A b & 18.87 & 143.69 & 3.95 & 1464 & 1.29 & 6170 & 1 & transit \\ 
\hline
31 & HAT-P-8 b & 14.50 & 405.33 & 3.08 & 1687 & 1.19 & 6200 & 1 & occultation \\ 
\hline
32 & WASP-54 b & 18.14 & 202.19 & 3.69 & 1531 & 1.15 & 6100 & 1 & transit \\ 
\hline
33 & WASP-109 b & 15.83 & 289.30 & 3.32 & 1729 & 0.91 & 6520 & 1 & transit \\ 
\hline
34 & HAT-P-41 b & 18.49 & 254.33 & 2.69 & 1886 & 1.42 & 6390 & 1 & transit \\ 
\hline
35 & HAT-P-13 b & 14.05 & 270.22 & 2.92 & 1600 & 1.22 & 5638 & 1 & transit \\ 
\hline
36 & KELT-15 b & 15.83 & 289.30 & 3.33 & 1500 & 1.18 & 6003 & 2 & transit \\ 
\hline
37 & KELT-7 b & 16.82 & 406.92 & 2.73 & 1996 & 1.53 & 6789 & 1 & transit \\ 
\hline
38 & HAT-P-6 b & 14.59 & 336.03 & 3.85 & 1629 & 1.29 & 6570 & 2 & transit \\ 
\hline
39 & WASP-49 b & 12.24 & 120.17 & 2.78 & 1334 & 0.94 & 5600 & 1 & transit \\ 
\hline
40 & WASP-15 b & 15.67 & 172.31 & 3.75 & 1609 & 1.18 & 6300 & 1 & transit \\ 
\hline
41 & WASP-79 b & 18.65 & 286.12 & 3.66 & 1709 & 1.56 & 6600 & 1 & transit \\ 
\hline
42 & KELT-4A b & 18.64 & 286.75 & 2.99 & 1779 & 1.20 & 6206 & 1 & transit \\ 
\hline
43 & WASP-17 b & 21.85 & 154.50 & 3.74 & 1725 & 1.31 & 6650 & 1 & transit \\ 
\hline
44 & WASP-3 b & 15.96 & 654.89 & 1.85 & 1933 & 1.24 & 6400 & 1 & occultation \\ 
\hline
45 & WASP-7 b & 14.59 & 305.19 & 4.95 & 1448 & 1.28 & 6400 & 1 & transit \\ 
\hline
46 & KELT-8 b & 20.41 & 275.63 & 3.24 & 1633 & 1.21 & 5754 & 1 & transit \\ 
\hline
47 & HAT-P-22 b & 11.85 & 682.55 & 3.21 & 1248 & 0.92 & 5302 & 3 & transit \\ 
\hline
48 & WASP-13 b & 15.44 & 158.95 & 4.35 & 1494 & 1.19 & 5989 & 1 & transit \\ 
\hline
49 & HAT-P-33 b & 20.05 & 242.56 & 3.47 & 1799 & 1.40 & 6446 & 1 & transit \\ 
\hline
50 & TrES-4 A b & 20.17 & 157.05 & 3.55 & 1644 & 1.45 & 6295 & 1 & transit \\ 
\hline
\end{tabular}%
}
\end{small}
\end{table}

\begin{table}[!ht]
\centering

\begin{small}
\resizebox{\columnwidth}{!}{%
\begin{tabular}{|c|c|c|c|c|c|c|c|c|c|c|}
\hline
\multirow{2}{*}{\#} & \multirow{2}{*}{\textbf{Planet}} &
\multicolumn{4}{c|}{\textbf{planetary properties}} &
\multicolumn{2}{c|}{\textbf{stellar properties}} &
\multicolumn{2}{c|}{\textbf{Observation}}\\
\cline{3-10}
  &  & { R $(R_{\oplus})$} &{\ M $(M_{\oplus})$} & {P (days)} & {T (K)} & { R $(R_{\odot}$)} &{ T (K)} & {\#} & {type} \\
\hline
51 & WASP-82 b & 18.33 & 394.20 & 2.71 & 2127 & 1.63 & 6490 & 1 & transit \\ 
\hline
52 & WASP-31 b & 16.87 & 151.96 & 3.41 & 1502 & 1.16 & 6200 & 1 & transit \\ 
\hline
53 & HAT-P-45 b & 15.65 & 283.57 & 3.13 & 1605 & 1.26 & 6330 & 2 & transit \\ 
\hline
54 & KELT-2A b & 14.33 & 472.41 & 4.11 & 1671 & 1.31 & 6148 & 1 & transit \\ 
\hline
55 & WASP-26 b & 14.06 & 326.81 & 2.76 & 1618 & 1.12 & 5950 & 2 & occultation \\ 
\hline
56 & TrES-2 & 12.83 & 398.34 & 2.47 & 1458 & 0.98 & 5850 & 4 & transit \\ 
\hline
57 & WASP-50 b & 12.62 & 467.32 & 1.96 & 1354 & 0.89 & 5400 & 5 & transit \\ 
\hline
58 & WASP-63 b & 15.69 & 120.80 & 4.38 & 1496 & 1.32 & 5570 & 1 & transit \\ 
\hline
59 & XO-2N b & 10.68 & 189.79 & 2.62 & 1312 & 0.97 & 5332 & 2 & transit \\ 
\hline
60 & WASP-104 b & 12.48 & 404.38 & 1.76 & 1476 & 1.08 & 5475 & 3 & occultation \\ 
\hline
61 & WASP-41 b & 13.28 & 292.47 & 3.05 & 1278 & 0.95 & 5450 & 2 & transit \\ 
\hline
62 & HAT-P-40 b & 18.98 & 195.51 & 4.46 & 1719 & 1.51 & 6080 & 1 & transit \\ 
\hline
63 & WASP-48 b & 18.33 & 311.55 & 2.14 & 1980 & 1.19 & 5920 & 1 & occultation \\ 
\hline
64 & HAT-P-4 b & 13.94 & 216.18 & 3.06 & 1653 & 1.26 & 5890 & 2 & transit \\ 
\hline
65 & WASP-4 b & 14.96 & 356.53 & 1.34 & 1818 & 0.93 & 5500 & 1 & occultation \\ 
\hline
66 & WASP-103 b & 17.59 & 467.32 & 0.93 & 2430 & 1.20 & 6110 & 1 & occultation \\ 
\hline
67 & WASP-75 b & 13.94 & 340.16 & 2.48 & 1660 & 1.14 & 6100 & 2 & occultation \\ 
\hline
68 & Qatar-1 b & 12.77 & 422.82 & 1.42 & 1347 & 0.85 & 4910 & 5 & occultation \\ 
\hline
69 & WASP-20 b & 16.00 & 99.50 & 4.90 & 1345 & 1.20 & 5950 & 1 & transit \\ 
\hline
70 & TrES-3 b & 14.32 & 607.20 & 1.31 & 1654 & 0.88 & 5720 & 1 & occultation \\ 
\hline
71 & PTFO 8-8695 b & 20.96 & 953.72 & 0.45 & 1884 & 0.34 & 3470 & 1 & occultation \\ 
\hline
72 & HAT-P-1 b & 13.35 & 166.58 & 4.47 & 1259 & 1.13 & 5975 & 1 & transit \\ 
\hline
73 & WASP-90 b & 17.89 & 200.28 & 3.92 & 1791 & 1.55 & 6430 & 2 & transit \\ 
\hline
74 & HAT-P-46 b & 14.09 & 156.73 & 4.47 & 1413 & 1.28 & 6120 & 1 & transit \\ 
\hline
75 & WASP-111 b & 15.82 & 581.77 & 2.31 & 2065 & 1.50 & 6400 & 1 & occultation \\ 
\hline
76 & XO-1 b & 12.99 & 286.12 & 3.94 & 1216 & 1.00 & 5940 & 1 & transit \\ 
\hline
77 & WASP-34 b & 13.39 & 187.57 & 4.32 & 1131 & 1.01 & 5700 & 1 & transit \\ 
\hline
78 & WASP-88 b & 18.65 & 178.03 & 4.95 & 1716 & 1.45 & 6431 & 1 & transit \\ 
\hline
79 & HATS-3 b & 18.62 & 361.78 & 3.55 & 1757 & 1.30 & 6351 & 2 & occultation \\ 
\hline
80 & WASP-100 b & 18.54 & 645.35 & 2.85 & 2143 & 1.57 & 6900 & 1 & occultation \\ 
\hline
81 & WASP-68 b & 13.61 & 302.01 & 5.08 & 1447 & 1.24 & 5911 & 2 & transit \\ 
\hline
82 & CoRoT-2 b & 16.08 & 1052.27 & 1.74 & 1484 & 0.97 & 5575 & 1 & occultation \\ 
\hline
83 & HAT-P-49 b & 15.51 & 549.98 & 2.69 & 2072 & 1.54 & 6820 & 1 & occultation \\ 
\hline
84 & HAT-P-56 b & 16.09 & 693.04 & 2.79 & 1791 & 1.30 & 6566 & 1 & occultation \\ 
\hline
85 & HAT-P-7 b & 16.01 & 543.30 & 2.20 & 2141 & 1.59 & 6310 & 1 & occultation \\ 
\hline
86 & WASP-21 b & 12.75 & 87.74 & 4.32 & 1298 & 0.89 & 5800 & 1 & transit \\ 
\hline
87 & WASP-22 b & 12.71 & 186.93 & 3.53 & 1383 & 1.10 & 6000 & 2 & transit \\ 
\hline
88 & WASP-24 b & 12.11 & 328.08 & 2.34 & 1611 & 1.13 & 6075 & 4 & occultation \\ 
\hline
89 & WASP-25 b & 13.83 & 184.39 & 3.76 & 1209 & 1.00 & 5750 & 1 & transit \\ 
\hline
90 & HAT-P-5 b & 13.74 & 336.98 & 2.79 & 1477 & 1.16 & 5960 & 6 & transit \\ 
\hline
91 & WASP-69 b & 11.60 & 82.66 & 3.87 & 938 & 0.83 & 4715 & 1 & transit \\ 
\hline
92 & WASP-87 b & 15.20 & 693.04 & 1.68 & 2251 & 1.20 & 6450 & 1 & occultation \\ 
\hline
93 & HAT-P-24 b & 13.63 & 217.77 & 3.36 & 1581 & 1.19 & 6329 & 4 & transit \\ 
\hline
94 & HAT-P-39 b & 17.24 & 190.43 & 3.54 & 1705 & 1.40 & 6430 & 3 & transit \\ 
\hline
95 & WASP-16 b & 11.06 & 271.81 & 3.12 & 1235 & 1.02 & 5550 & 4 & transit \\ 
\hline
96 & TrES-1 b & 12.06 & 241.93 & 3.03 & 1147 & 0.88 & 5250 & 2 & transit \\ 
\hline
97 & WASP-64 b & 13.95 & 404.06 & 1.57 & 1587 & 0.98 & 5400 & 3 & occultation \\ 
\hline
98 & WASP-6 b & 13.43 & 159.91 & 3.36 & 1161 & 0.89 & 5450 & 1 & transit \\ 
\hline
99 & WASP-55 b & 14.27 & 181.21 & 4.47 & 1236 & 1.01 & 5900 & 1 & transit \\ 
\hline
100 & HAT-P-36 b & 13.87 & 582.41 & 1.33 & 1778 & 1.02 & 5580 & 1 & occultation \\ 
\hline
\end{tabular}%
}
\end{small}
\end{table}

\begin{table}[!ht]
\centering

\begin{small}
\resizebox{\columnwidth}{!}{%
\begin{tabular}{|c|c|c|c|c|c|c|c|c|c|c|}
\hline
\multirow{2}{*}{\#} & \multirow{2}{*}{\textbf{Planet}} &
\multicolumn{4}{c|}{\textbf{planetary properties}} &
\multicolumn{2}{c|}{\textbf{stellar properties}} &
\multicolumn{2}{c|}{\textbf{Observation}}\\
\cline{3-10}
  &  & { R $(R_{\oplus})$} &{\ M $(M_{\oplus})$} & {P (days)} & {T (K)} & { R $(R_{\odot}$)} &{ T (K)} & {\#} & {type} \\
\hline
101 & HAT-P-9 b & 15.36 & 213.00 & 3.92 & 1490 & 1.28 & 6350 & 4 & transit \\ 
\hline
102 & HAT-P-14 b & 13.17 & 699.40 & 4.63 & 1525 & 1.39 & 6600 & 3 & occultation \\ 
\hline
103 & WASP-28 b & 13.31 & 288.34 & 3.41 & 1429 & 1.02 & 6150 & 6 & transit \\ 
\hline
104 & XO-4 b & 14.70 & 546.80 & 4.13 & 1418 & 1.32 & 5700 & 6 & occultation \\ 
\hline
105 & WASP-58 b & 15.03 & 282.94 & 5.02 & 1242 & 0.94 & 5800 & 2 & transit \\ 
\hline
106 & HAT-P-23 b & 15.01 & 664.43 & 1.21 & 1997 & 1.13 & 5905 & 1 & occultation \\ 
\hline
107 & Qatar-2 b & 12.55 & 790.63 & 1.34 & 1256 & 0.74 & 4645 & 9 & occultation \\ 
\hline
108 & WASP-5 b & 12.85 & 520.41 & 1.63 & 1693 & 1.00 & 5700 & 2 & occultation \\ 
\hline
109 & WASP-65 b & 12.20 & 492.76 & 2.31 & 1446 & 0.93 & 5600 & 7 & occultation \\ 
\hline
110 & CoRoT-1 b & 16.35 & 327.44 & 1.51 & 1839 & 0.95 & 6298 & 1 & occultation \\ 
\hline
111 & HAT-P-27 b & 11.19 & 197.10 & 3.04 & 1161 & 0.92 & 5300 & 3 & transit \\ 
\hline
112 & KELT-6 b & 12.95 & 140.51 & 7.85 & 1284 & 1.13 & 6272 & 1 & transit \\ 
\hline
113 & WASP-45 b & 12.73 & 320.13 & 3.13 & 1165 & 0.91 & 5140 & 6 & transit \\ 
\hline
114 & WASP-72 b & 11.08 & 448.25 & 2.22 & 1819 & 1.23 & 6250 & 2 & occultation \\ 
\hline
115 & HATS-1 b & 14.29 & 589.72 & 3.45 & 1332 & 0.99 & 5870 & 14 & transit \\ 
\hline
116 & WASP-78 b & 19.20 & 368.77 & 2.18 & 2136 & 2.02 & 6100 & 1 & occultation \\ 
\hline
117 & WASP-96 b & 13.17 & 152.60 & 3.43 & 1251 & 1.06 & 5540 & 3 & transit \\ 
\hline
118 & HAT-P-28 b & 13.30 & 199.01 & 3.26 & 1345 & 1.02 & 5680 & 6 & transit \\ 
\hline
119 & WASP-39 b & 13.94 & 89.01 & 4.06 & 1088 & 0.93 & 5400 & 1 & transit \\ 
\hline
120 & WASP-80 b & 10.45 & 176.12 & 3.07 & 794 & 0.57 & 4145 & 1 & transit \\ 
\hline
121 & HATS-2 b & 12.82 & 427.58 & 1.35 & 1528 & 0.88 & 5227 & 5 & occultation \\ 
\hline
122 & WASP-71 b & 16.02 & 712.75 & 2.90 & 1987 & 1.56 & 6050 & 1 & occultation \\ 
\hline
123 & WASP-38 b & 11.96 & 861.53 & 6.87 & 1218 & 1.23 & 6150 & 10 & transit \\ 
\hline
124 & WASP-110 b & 13.58 & 162.13 & 3.78 & 1113 & 0.89 & 5400 & 1 & transit \\ 
\hline
125 & HAT-P-3 b & 9.07 & 187.88 & 2.90 & 1115 & 0.92 & 5224 & 4 & transit \\ 
\hline
126 & WASP-47 b & 12.84 & 336.98 & 4.16 & 1240 & 1.04 & 5576 & 8 & transit \\ 
\hline
127 & WASP-98 b & 12.07 & 263.86 & 2.96 & 1149 & 0.69 & 5525 & 8 & transit \\ 
\hline
128 & WASP-46 b & 14.38 & 667.92 & 1.43 & 1615 & 0.96 & 5620 & 3 & occultation \\ 
\hline
129 & HAT-P-25 b & 13.06 & 180.22 & 3.65 & 1172 & 1.01 & 5500 & 3 & transit \\ 
\hline
130 & WASP-18 b & 12.78 & 3315.77 & 0.94 & 2345 & 1.24 & 6400 & 1 & occultation \\ 
\hline
131 & WASP-67 b & 15.36 & 133.52 & 4.61 & 1000 & 0.87 & 5200 & 1 & transit \\ 
\hline
132 & WASP-14 b & 14.06 & 2333.75 & 2.24 & 1834 & 1.21 & 6462 & 1 & occultation \\ 
\hline
133 & WASP-60 b & 9.44 & 163.40 & 4.31 & 1261 & 0.51 & 5900 & 6 & transit \\ 
\hline
134 & WASP-11 b & 11.47 & 146.24 & 3.72 & 1002 & 0.82 & 4974 & 1 & transit \\ 
\hline
135 & HAT-P-35 b & 14.62 & 335.07 & 3.65 & 1537 & 1.24 & 6096 & 12 & occultation \\ 
\hline
136 & WASP-36 b & 13.93 & 724.51 & 1.54 & 1655 & 1.02 & 5881 & 3 & occultation \\ 
\hline
137 & HAT-P-50 b & 14.13 & 429.17 & 3.12 & 1805 & 1.27 & 6280 & 3 & occultation \\ 
\hline
138 & WASP-99 b & 12.07 & 883.78 & 5.75 & 1438 & 1.48 & 6180 & 6 & occultation \\ 
\hline
139 & HAT-P-42 b & 14.01 & 309.96 & 4.64 & 1389 & 1.18 & 5743 & 13 & transit \\ 
\hline
140 & WASP-73 b & 12.73 & 597.66 & 4.09 & 1736 & 1.34 & 6036 & 3 & occultation \\ 
\hline
141 & WASP-135 b & 14.27 & 604.02 & 1.40 & 1673 & 0.98 & 5675 & 3 & occultation \\ 
\hline
142 & WASP-23 b & 10.56 & 281.03 & 2.94 & 1099 & 0.78 & 5150 & 10 & transit \\ 
\hline
143 & TrES-5 b & 13.27 & 565.24 & 1.48 & 1433 & 0.88 & 5171 & 10 & occultation \\ 
\hline
144 & HAT-P-16 b & 13.06 & 1332.98 & 2.78 & 1527 & 1.22 & 6140 & 3 & occultation \\ 
\hline
145 & Kepler-12 b & 19.20 & 136.70 & 4.44 & 1341 & 1.09 & 5953 & 2 & transit \\ 
\hline
146 & Kepler-7 b & 17.71 & 137.65 & 4.89 & 1584 & 1.36 & 5933 & 4 & transit \\ 
\hline
147 & WASP-44 b & 11.00 & 276.26 & 2.42 & 1275 & 0.92 & 5410 & 28 & transit \\ 
\hline
148 & XO-5 b & 11.30 & 342.39 & 4.19 & 1206 & 0.88 & 5510 & 16 & transit \\ 
\hline
149 & HAT-P-43 b & 14.08 & 209.82 & 3.33 & 1322 & 1.05 & 5645 & 12 & transit \\ 
\hline
150 & HAT-P-55 b & 12.97 & 185.02 & 3.58 & 1278 & 1.01 & 5808 & 10 & transit \\ 
\hline
\end{tabular}%
}
\end{small}
\end{table}

\begin{table}[!ht]
\centering

\begin{small}
\resizebox{\columnwidth}{!}{%
\begin{tabular}{|c|c|c|c|c|c|c|c|c|c|c|}
\hline
\multirow{2}{*}{\#} & \multirow{2}{*}{\textbf{Planet}} &
\multicolumn{4}{c|}{\textbf{planetary properties}} &
\multicolumn{2}{c|}{\textbf{stellar properties}} &
\multicolumn{2}{c|}{\textbf{Observation}}\\
\cline{3-10}
  &  & { R $(R_{\oplus})$} &{\ M $(M_{\oplus})$} & {P (days)} & {T (K)} & { R $(R_{\odot}$)} &{ T (K)} & {\#} & {type} \\
\hline
151 & WASP-32 b & 13.06 & 1144.46 & 2.72 & 1507 & 1.10 & 6100 & 5 & occultation \\ 
\hline
152 & HAT-P-29 b & 12.15 & 247.33 & 5.72 & 1224 & 1.21 & 6087 & 8 & transit \\ 
\hline
153 & WASP-10 b & 11.85 & 972.79 & 3.09 & 1009 & 0.71 & 4675 & 37 & transit \\ 
\hline
154 & Kepler-6 b & 14.27 & 213.00 & 3.23 & 1354 & 1.05 & 5640 & 15 & transit \\ 
\hline
155 & HD219134b & 1.61 & 4.47 & 3.09 & 934 & 0.78 & 4699 & 1 & transit \\ 
\hline
156 & HATS-13 b & 13.30 & 172.62 & 3.04 & 1212 & 0.96 & 5523 & 10 & transit \\ 
\hline
157 & HAT-P-51 b & 14.19 & 98.23 & 4.22 & 1159 & 0.98 & 5449 & 2 & transit \\ 
\hline
158 & HAT-P-34 b & 13.14 & 1057.99 & 5.45 & 1440 & 1.39 & 6442 & 10 & occultation \\ 
\hline
159 & WASP-37 b & 12.47 & 539.17 & 3.58 & 1293 & 0.85 & 5800 & 43 & transit \\ 
\hline
160 & WASP-56 b & 11.98 & 181.52 & 4.62 & 1117 & 1.03 & 5600 & 4 & transit \\ 
\hline
161 & WASP-66 b & 15.25 & 737.54 & 4.09 & 1754 & 1.30 & 6600 & 3 & occultation \\ 
\hline
162 & WASP-112 b & 13.07 & 279.76 & 3.04 & 1349 & 0.81 & 5610 & 28 & transit \\ 
\hline
163 & HAT-P-44 b & 14.05 & 124.62 & 4.30 & 1092 & 0.94 & 5295 & 2 & transit \\ 
\hline
164 & HAT-P-37 b & 12.93 & 371.63 & 2.80 & 1166 & 0.93 & 5500 & 37 & transit \\ 
\hline
165 & Gliese 436 b & 4.13 & 23.11 & 2.64 & 695 & 0.45 & 3684 & 1 & transit \\ 
\hline
166 & WASP-29 b & 8.69 & 77.57 & 3.92 & 970 & 0.82 & 4800 & 1 & transit \\ 
\hline
167 & HD 219134 b & 1.57 & 3.81 & 3.09 & 931 & 0.79 & 4699 & 1 & transit \\ 
\hline
168 & HAT-P-12 b & 10.52 & 67.08 & 3.21 & 932 & 0.73 & 4650 & 1 & transit \\ 
\hline
169 & Kepler-13 A b & 15.43 & 2571.87 & 1.76 & 2389 & 1.72 & 7200 & 1 & occultation \\ 
\hline
170 & HAT-P-19 b & 12.17 & 92.83 & 4.01 & 982 & 0.84 & 4990 & 1 & transit \\ 
\hline
171 & CoRoT-11 b & 15.25 & 791.59 & 2.99 & 1686 & 1.27 & 6440 & 5 & occultation \\ 
\hline
172 & Kepler-8 b & 15.58 & 187.57 & 3.52 & 1528 & 1.13 & 6251 & 20 & transit \\ 
\hline
173 & HATS-10 b & 10.63 & 167.22 & 3.31 & 1369 & 1.10 & 5880 & 28 & transit \\ 
\hline
174 & WTS-2 b & 14.96 & 356.06 & 1.02 & 1495 & 0.82 & 5000 & 16 & occultation \\ 
\hline
175 & HAT-P-52 b & 11.07 & 260.05 & 2.75 & 1184 & 0.89 & 5131 & 43 & transit \\ 
\hline
176 & HAT-P-20 b & 9.51 & 2303.55 & 2.88 & 946 & 0.76 & 4595 & 97 & occultation \\ 
\hline
177 & WASP-120 b & 16.62 & 1592.71 & 3.61 & 1842 & 1.45 & 6450 & 1 & occultation \\ 
\hline
178 & HATS-9 b & 11.69 & 266.09 & 1.92 & 1769 & 1.03 & 5366 & 9 & occultation \\ 
\hline
179 & CoRoT-19 b & 15.91 & 352.88 & 3.90 & 1616 & 1.21 & 6090 & 16 & occultation \\ 
\hline
180 & OGLE-TR-10 b & 18.87 & 216.18 & 3.10 & 1554 & 1.28 & 6075 & 16 & occultation \\ 
\hline
181 & WASP-42 b & 11.85 & 158.95 & 4.98 & 969 & 0.88 & 5200 & 1 & transit \\ 
\hline
182 & WASP-61 b & 13.61 & 654.89 & 3.86 & 1509 & 1.22 & 6250 & 17 & occultation \\ 
\hline
183 & HAT-P-31 b & 11.74 & 690.18 & 5.01 & 1343 & 1.22 & 6065 & 43 & occultation \\ 
\hline
184 & HAT-P-53 b & 14.46 & 471.77 & 1.96 & 1624 & 1.09 & 5956 & 7 & occultation \\ 
\hline
185 & WASP-8 b & 11.39 & 713.38 & 8.16 & 906 & 1.03 & 5600 & 6 & transit \\ 
\hline
186 & HATS-4 b & 11.19 & 420.59 & 2.52 & 1282 & 1.00 & 5403 & 97 & occultation \\ 
\hline
187 & Kepler-447 b & 18.11 & 435.53 & 7.79 & 908 & 0.76 & 5493 & 2 & transit \\ 
\hline
188 & Kepler-76 b & 14.92 & 638.99 & 1.54 & 2074 & 1.20 & 6409 & 2 & occultation \\ 
\hline
189 & WASP-57 b & 10.05 & 213.63 & 2.84 & 1430 & 1.01 & 5600 & 43 & occultation \\ 
\hline
190 & CoRoT-5 b & 15.23 & 148.46 & 4.04 & 1315 & 1.00 & 6100 & 15 & transit \\ 
\hline
191 & HD 17156 b & 12.02 & 1014.44 & 21.22 & 816 & 1.27 & 6079 & 9 & transit \\ 
\hline
192 & Kepler-412 b & 14.54 & 298.51 & 1.72 & 1780 & 1.17 & 5750 & 12 & occultation \\ 
\hline
193 & XO-3 b & 13.35 & 3748.12 & 3.19 & 1665 & 1.21 & 6429 & 1 & occultation \\ 
\hline
194 & WASP-117 b & 11.20 & 87.58 & 10.02 & 997 & 1.13 & 6040 & 1 & transit \\ 
\hline
195 & Gliese 1214 b & 2.77 & 6.20 & 1.58 & 552 & 0.18 & 3250 & 1 & transit \\ 
\hline
196 & Gliese 3470 b & 3.80 & 13.73 & 3.34 & 635 & 0.51 & 3652 & 1 & transit \\ 
\hline
197 & HAT-P-11 b & 4.96 & 25.75 & 4.89 & 848 & 0.81 & 4780 & 1 & transit \\ 
\hline
198 & GJ 1132 b & 1.16 & 1.62 & 1.63 & 529 & 0.18 & 3270 & 1 & transit \\ 
\hline
199 & Gliese 436 c & 0.66 & 0.28 & 1.37 & 813 & 0.45 & 3684 & 1 & transit \\ 
\hline
200 & HAT-P-26 b & 6.20 & 18.76 & 4.23 & 967 & 0.82 & 5079 & 1 & transit \\ 
\hline
201 & HAT-P-18 b & 10.39 & 62.31 & 5.51 & 818 & 0.77 & 4870 & 1 & transit \\ 
\hline
202 & HD 97658 b & 2.34 & 7.55 & 9.49 & 729 & 0.77 & 5119 & 1 & transit \\ 
\hline
203 & HAT-P-17 b & 11.08 & 169.76 & 10.34 & 758 & 0.86 & 5246 & 1 & transit \\ 
\hline
204 & WASP-84 b & 10.70 & 222.53 & 8.52 & 780 & 0.85 & 5280 & 1 & transit \\ 
\hline
205 & HATS-6 b & 10.95 & 101.41 & 3.33 & 693 & 0.57 & 3724 & 1 & transit \\ 
\hline
206 & \footnotesize{EPIC 203771098 c} & 7.93 & 27.02 & 42.36 & 596 & 1.12 & 5743 & 1 & transit \\ 
\hline
207 & KOI-142 b & 4.13 & 1.76 & 10.95 & 764 & 1.02 & 5513 & 1 & transit \\ 
\hline
208 & HATS-5 b & 10.01 & 75.34 & 4.76 & 998 & 0.94 & 5304 & 1 & transit \\ 
\hline
209 & Kepler-51 b & 6.95 & 2.10 & 45.15 & 496 & 1.04 & 6018 & 1 & transit \\ 
\hline
210 & HAT-P-2 b & 10.44 & 2778.51 & 5.63 & 1443 & 1.36 & 6290 & 1 & occultation \\ 
\hline
 
\end{tabular}%
}
\end{small}
\caption{List of known planets observable by ARIEL. The former to last column represents the number of transits/eclipses necessary to fulfil the ARIEL Tier 1 goals.}     
\label{tab:all_known_planets}
\end{table}

\end{appendices}

\clearpage

%REFERENCES
\bibliographystyle{apalike}
\bibliography{targetlist}

\begin{thebibliography}{}

\bibitem[{Baraffe} et~al., 1998]{1998A&A...337..403B}
{Baraffe}, I., {Chabrier}, G., {Allard}, F., and {Hauschildt}, P.~H. (1998).
\newblock {Evolutionary models for solar metallicity low-mass stars:
  mass-magnitude relationships and color-magnitude diagrams}.
\newblock {\em \aap}, 337:403--412.

\bibitem[{Casagrande} et~al., 2011]{2011A&A...530A.138C}
{Casagrande}, L., {Sch{\"o}nrich}, R., {Asplund}, M., {Cassisi}, S.,
  {Ram{\'{\i}}rez}, I., {Mel{\'e}ndez}, J., {Bensby}, T., and {Feltzing}, S.
  (2011).
\newblock {New constraints on the chemical evolution of the solar neighbourhood
  and Galactic disc(s). Improved astrophysical parameters for the
  Geneva-Copenhagen Survey}.
\newblock {\em \aap}, 530:A138.

\bibitem[{Chen} and {Kipping}, 2016]{2016arXiv160308614C}
{Chen}, J. and {Kipping}, D.~M. (2016).
\newblock {Probabilistic Forecasting of the Masses and Radii of Other Worlds}.
\newblock {\em ArXiv e-prints}.

\bibitem[{Cohen} et~al., 2003]{2003AJ....126.1090C}
{Cohen}, M., {Wheaton}, W.~A., and {Megeath}, S.~T. (2003).
\newblock {Spectral Irradiance Calibration in the Infrared. XIV. The Absolute
  Calibration of 2MASS}.
\newblock {\em \aj}, 126:1090--1096.

\bibitem[{Fressin} et~al., 2013]{2013ApJ...766...81F}
{Fressin}, F., {Torres}, G., {Charbonneau}, D., {Bryson}, S.~T.,
  {Christiansen}, J., {Dressing}, C.~D., {Jenkins}, J.~M., {Walkowicz}, L.~M.,
  and {Batalha}, N.~M. (2013).
\newblock {The False Positive Rate of Kepler and the Occurrence of Planets}.
\newblock {\em \apj}, 766:81.

\bibitem[{Mulders} et~al., 2015]{2015ApJ...814..130M}
{Mulders}, G.~D., {Pascucci}, I., and {Apai}, D. (2015).
\newblock {An Increase in the Mass of Planetary Systems around Lower-mass
  Stars}.
\newblock {\em \apj}, 814:130.

\bibitem[{Mulders} et~al., 2016]{2016arXiv160905898M}
{Mulders}, G.~D., {Pascucci}, I., {Apai}, D., {Frasca}, A., and
  {Molenda-Zakowicz}, J. (2016).
\newblock {A Super-Solar Metallicity For Stars With Hot Rocky Exoplanets}.
\newblock {\em ArXiv e-prints}.

\bibitem[{{\"O}berg} et~al., 2011]{2011ApJ...740..109O}
{{\"O}berg}, K.~I., {Boogert}, A.~C.~A., {Pontoppidan}, K.~M., {van den Broek},
  S., {van Dishoeck}, E.~F., {Bottinelli}, S., {Blake}, G.~A., and {Evans}, II,
  N.~J. (2011).
\newblock {The Spitzer Ice Legacy: Ice Evolution from Cores to Protostars}.
\newblock {\em \apj}, 740:109.

\bibitem[{Pascale} et~al., 2015]{2015ExA....40..601P}
{Pascale}, E., {Waldmann}, I.~P., {MacTavish}, C.~J., {Papageorgiou}, A.,
  {Amaral-Rogers}, A., {Varley}, R., {Coud{\'e} du Foresto}, V., {Griffin},
  M.~J., {Ollivier}, M., {Sarkar}, S., {Spencer}, L., {Swinyard}, B.~M.,
  {Tessenyi}, M., and {Tinetti}, G. (2015).
\newblock {EChOSim: The Exoplanet Characterisation Observatory software
  simulator}.
\newblock {\em Experimental Astronomy}, 40:601--619.

\bibitem[{Piskorz} et~al., 2015]{2015ApJ...814..148P}
{Piskorz}, D., {Knutson}, H.~A., {Ngo}, H., {Muirhead}, P.~S., {Batygin}, K.,
  {Crepp}, J.~R., {Hinkley}, S., and {Morton}, T.~D. (2015).
\newblock {Friends of Hot Jupiters. III. An Infrared Spectroscopic Search for
  Low-mass Stellar Companions}.
\newblock {\em \apj}, 814:148.

\bibitem[{Puig} et~al., 2015]{ESA...Rad...Mod}
{Puig}, L., {Isaak}, K., {Linder}, M., {Escudero}, I., {Crouzet}, P.-E.,
  {Walker}, R., {Ehle}, M., {H{\"u}bner}, J., {Timm}, R., {de Vogeleer}, B.,
  {Drossart}, P., {Hartogh}, P., {Lovis}, C., {Micela}, G., {Ollivier}, M.,
  {Ribas}, I., {Snellen}, I., {Swinyard}, B., {Tinetti}, G., and {Eccleston},
  P. (2015).
\newblock {The phase 0/A study of the ESA M3 mission candidate EChO}.
\newblock {\em Experimental Astronomy}, 40:393--425.

\bibitem[{Ribas} and {Lovis}, 2013]{ESA...EChO...MRS}
{Ribas}, I. and {Lovis}, C. (2013).
\newblock Echo targets: the mission reference sample and beyond
  [echo-sre-sa-phasea-001\_mrsv2.4], european space agency.

\bibitem[{Sarkar} et~al., 2016]{2016SPIE.9904E..3RS}
{Sarkar}, S., {Papageorgiou}, A., and {Pascale}, E. (2016).
\newblock {Exploring the potential of the ExoSim simulator for transit
  spectroscopy noise estimation}.
\newblock In {\em Space Telescopes and Instrumentation 2016: Optical, Infrared,
  and Millimeter Wave}, volume 9904, page 99043R.

\bibitem[{Sarkar} and {Pascale}, 2015]{2015EPSC...10..187S}
{Sarkar}, S. and {Pascale}, E. (2015).
\newblock {ExoSim: a novel simulator of exoplanet spectroscopic observations}.
\newblock {\em European Planetary Science Congress 2015, held 27 September - 2
  October, 2015 in Nantes, France, Online at <A
  href=``http://meetingorganizer.copernicus.org/EPSC2015/EPSC2015''>
  http://meetingorganizer.copernicus.org/EPSC2015</A>, id.EPSC2015-187},
  10:EPSC2015--187.

\bibitem[{Sullivan} et~al., 2015]{2015ApJ...809...77S}
{Sullivan}, P.~W., {Winn}, J.~N., {Berta-Thompson}, Z.~K., {Charbonneau}, D.,
  {Deming}, D., {Dressing}, C.~D., {Latham}, D.~W., {Levine}, A.~M.,
  {McCullough}, P.~R., {Morton}, T., {Ricker}, G.~R., {Vanderspek}, R., and
  {Woods}, D. (2015).
\newblock {The Transiting Exoplanet Survey Satellite: Simulations of Planet
  Detections and Astrophysical False Positives}.
\newblock {\em \apj}, 809:77.

\bibitem[{Triaud} et~al., 2014]{2014MNRAS.444..711T}
{Triaud}, A.~H.~M.~J., {Lanotte}, A.~A., {Smalley}, B., and {Gillon}, M.
  (2014).
\newblock {Colour-magnitude diagrams of transiting Exoplanets - II. A larger
  sample from photometric distances}.
\newblock {\em \mnras}, 444:711--728.

\bibitem[{Valencia} et~al., 2007]{2007ApJ...665.1413V}
{Valencia}, D., {Sasselov}, D.~D., and {O'Connell}, R.~J. (2007).
\newblock {Detailed Models of Super-Earths: How Well Can We Infer Bulk
  Properties?}
\newblock {\em \apj}, 665:1413--1420.

\bibitem[{Venot} et~al., 2015]{2015A&A...577A..33V}
{Venot}, O., {H{\'e}brard}, E., {Ag{\'u}ndez}, M., {Decin}, L., and
  {Bounaceur}, R. (2015).
\newblock {New chemical scheme for studying carbon-rich exoplanet atmospheres}.
\newblock {\em \aap}, 577:A33.

\end{thebibliography}

%REFERENCES

\end{document}